\documentclass[final,12pt]{elsarticle}

\usepackage[margin=1in]{geometry}
\usepackage[ansinew]{inputenc}
\usepackage[english]{babel}
\usepackage{fancyhdr}
\usepackage[active]{srcltx}
\usepackage{amsmath,amssymb}
\usepackage{xcolor}
\usepackage{array}
\usepackage{textcomp}
\usepackage{tikz}
\usetikzlibrary{shapes.geometric, arrows,shadows}
\usepackage[ruled,vlined]{algorithm2e}
\usepackage{cuted}
\usepackage{placeins}
\usepackage[normalem]{ulem}
\usepackage{hyperref}
\usepackage{booktabs}

\usepackage{xcolor}
\usepackage{bm}
\usepackage{stackengine}
\usepackage{subfig}
\usepackage{smartdiagram}
\usesmartdiagramlibrary{additions}
\tikzstyle{myblock} = [rectangle,
        general shadow={%
            fill=gray!50,
            shadow xshift= 2pt,
            shadow yshift=-2pt
            },
            font = {\scriptsize},  rounded corners, minimum width=1.5cm, text width=1.5cm,minimum height=2.5cm,text centered,line width=.5pt, draw=black, top color=white, bottom color=black!10]%,fill=black!10]
\tikzstyle{feedback} = [rectangle,font = {\scriptsize},  general shadow={%
            fill=gray!50,
            shadow xshift= 2pt,
            shadow yshift=-2pt
            },
            rounded corners, minimum width=1.6cm, text width=4cm,minimum height=1cm,text centered, line width=.5pt,draw=black, top color=white, bottom color=black!10]%,fill=black!10]
\tikzstyle{arrow} = [line width=3pt,opacity=.4,draw=gray!50,->,>=stealth]%ultra thick,

\addto\extrasenglish{%
}

\journal{Mechanical Systems and Signal Processing}

\begin{document}

\begin{frontmatter}



\title{Iterative method for real-time Hybrid testing: application to a cantilever beam with two interface degrees of freedom}


\author{Alessandra~Vizzaccaro$^{a}$,
        Sandor~Beregi$^{b}$,
        David~A.W.~Barton$^{c}$,
        Simon~A.~Neild$^{d}$}

\affiliation[a]{
organization={College of Engineering, Mathematics and Physical Sciences, University of Exeter},city={Exeter},postcode={EX4 4QJ}, country={UK}
}
\affiliation[b]{
organization={Faculty of Medicine, School of Public Health, Imperial College London},city={London},postcode={SW7 2AZ}, country={UK}
}
\affiliation[c]{organization={Faculty of Engineering, University of Bristol},city={Bristol},postcode={BS8 1UB}, country={UK}
}

\begin{abstract}
{In this paper, an iterative method for real-time hybrid testing (RTHT) is proposed. The method seeks to iteratively balance the interface conditions between the physical and numerical substructures by controlling the periodic demand of the actuators. It is then suitable for RTHT of structures undergoing a periodic response, e.g. structures excited at resonance. We demonstrate the capabilities of the method on a cantilever beam in bending motion with two degrees of freedom at the interface, which we use as a prototype for future testing of aircraft wings. We show that a number of challenges arise in these settings, such as the difficulty in measuring interface forces while controlling a continuous structure and the instability of the hybrid test for small time delays. Classical RTHT strategies could produce inaccurate or unstable outcomes, whereas the proposed method is able to attain very good interface synchronisation in a wide range of tested scenarios.
%We show that classical RTHT is particularly challenging on such systems for a number of reason. First, to control the interface of a continuous structure the forces need to be measured at the interface, which conflict with the need of controlling its motion. Secondly, two degrees of freedom at the interface are required for a pure bending test. Lastly, the low damping and relatively high frequency of the structure render RTHT  Rig independent and control scheme independent. Suitable to resonant tests. No risk of instability which for the tested structure is high as shown in the substructurability study. Accuracy limited only by sensors. is challenging on continuous structure. Enable wing testing. On continuous structure it is hard to sense at the interface. Proposed SG measurement that combined with proposed method increases accuracy. Main challenges, 2 DOF, very low damping, continuous structure. Main result: rig to test cantilever structures. Great interface synchronisation attained for wide range of test. 
}
\end{abstract}


% \begin{highlights}
% \end{highlights}


\end{frontmatter}

\section{Introduction}
Real-time hybrid testing (RTHT) is an experimental technique for testing a critical part of a structure within the context of the whole assembly. It is sometimes referred to as hardware-in-the-loop, a cyber-physical system, or experimental dynamic substructuring. In RTHT, the original structure is split into two components, the physical substructure (PS) that is physically realised and a numerical substructure (NS) that is virtually simulated. The two together form the hybrid structure (HS) which should reproduce the behaviour of the original one. The method, originally developed in the context of earthquake engineering \cite{Hakuno,Nakashima}, has been applied to mechanical \cite{Blakeborough} and aeroelastic systems \cite{FagleyAero}. The appeal of this technique lies in its capacity to fully exploit the advantages of both numerical modelling and experimental testing at the same time. Usually, the physical substructure is a critical component, for example, where damage is expected, or more generally, one that is not understood well enough to be accurately modelled. At the same time, the numerical substructure replaces the part of the assembly that can be reliably modelled \cite{Carrion}. Large structures that do not fit in the experimental facility can then be tested under such a hybrid paradigm. Moreover, the parameters of the numerical substructure can easily be modified, and so reproduce different environments or explore a certain design space in which the critical component must be tested \cite{Ruffini}. 

In RTHT, the numerical and physical substructures are joined by an interface, where equilibrium of forces and compatibility of displacements must be attained. A control system comprising sensors, actuators, a transfer system (intended here as all the mechanical components connecting the actuators to the physical substructure), and a control algorithm determines the dynamics of the interface. Often the experiment is performed with displacement control or force control, but velocity or acceleration control is equally applicable. The reaction forces at the interface are then measured and sent to the simulator containing the numerical model of the NS where the reference target signal was computed. If the dynamics of the two substructures is perfectly synchronous in both force and displacement, then the response of the hybrid structure is identical to that of the original. However, this is only possible in the ideal case of a perfect unitary control system, which imposes the command signal on the physical substructure instantly and  without mismatches. In reality, the realisation of the command signal can only be achieved after a certain delay and attenuation, {which typically has frequency-dependent characteristics}. For the accuracy of the test to be acceptable, such a time delay must be significantly smaller than the characteristic time of the tested structure, meaning that the feedback loop internal to the control system has to be much faster than the outer one. This represents the main challenge of RTHT, especially when high-force actuators are needed, which can be slow, or when high frequencies of the structure are of interest \cite{Carrion}. Another source of delay is the time spent on the numerical model calculations; however, it is reasonable to consider it negligible compared to the overall control system delay.

Several contributions \cite{Wallace2005, Kyrychko2007, Terkovics2016, Maghareh2017} have studied the stability of RTHT for different structures, simplifying the dynamics of the control system to a pure time delay. Generally, the response of the actuators is frequency dependent, so the constant time delay assumption is a simplification. However, the stability boundaries can still give a very good indication of RTHT feasibility, i.e., what has been called substructurability \cite{Terkovics2016} of a given structure. A similar concept is proposed in \cite{Maghareh2017} where the so-called Predictive Stability Indicator is defined as the critical delay of an RTHT system at which stability is lost.  

Provided the hybrid structure is stable for a given interface delay, the success of the experiment is still not guaranteed. In fact, a mismatch between the command and the realised signal affects not only the stability but also the accuracy of the experiment. Since a model of the PS is typically not available, there is no reference solution of the original structure to assess the accuracy of the hybrid test results. Moreover, different hybrid structures have different sensitivities to errors at the interface. For example, a lightly-damped structure is much more sensitive to interface delays and lags than a heavily damped one, especially around resonance \cite{Mosqueda2007II}. The problem of how accurately the RTHT reproduces the behaviour of the original structure is the subject of several studies aimed at providing accuracy measures based on interface data. Examples of accuracy measures are the synchronisation subspace plot (SSP) proposed in \cite{Wallace2005} to visualise the amplitude and phase mismatch between the NS and PS sides of the interface, the hybrid simulation error monitor (HSEM) defined in \cite{Mosqueda2007II} as the energy introduced in the experiment by the control system, and the predictive performance indicator (PPI) introduced in \cite{Maghareh2014}, where adjustments to the interface location are also investigated to increase the fidelity of the response. 
The interested reader is referred to \cite{InsamReview,Christenson2014} for more detailed reviews of accuracy measures for RTHT.

Since the fidelity of an RTHT experiment depends on interface errors, a large body of literature on RTHT is devoted to the development of control schemes able to realise the coupling between NS and PS as accurately as possible. Moreover, since the main cause of inaccuracies is the delay introduced by the dynamics of the control system, most of the literature deals with the development of delay compensation techniques.
Several strategies have been proposed such as
polynomial extrapolation \cite{Horiuchi},  adaptive schemes \cite{Wagg2001,Wallace2005II}, feed-forward compensation based on the inverse model of the control system \cite{Chen2009}, and, more recently, techniques that exploit normalised passivity control \cite{PeirisPassivity}, robust model predictive control (MPC) \cite{Tsokanas}, and data-driven modelling of control systems \cite{SimpsonData}.

In this paper, we propose an alternative strategy that is particularly suitable for testing mechanical components at resonance. This is a common testing requirement, as it is often the critical loading case and highlights the damping of the system. Such tests result in periodic responses, a feature which is exploited in the proposed control method. The method, inspired by control-based continuation \cite{BartonCBC2010,BartonCBC2017}, is to control the interface by providing a periodic input to the actuators that is updated iteratively every few periods until the error across the interface is below a threshold level. In this way, the two substructures are completely decoupled in time, thus lifting the real-time constraint and making interface delays irrelevant. The fidelity of the RTHT experiment is then only limited by the accuracy of the measurement of the interface quantities. In the recent works of Witteveen \textit{et al.} \cite{Witteveen2022,Witteveen2023}, a similar approach is proposed called non-simultaneous real-time hybrid simulation, however the details of the experimental implementation are limited and the experimental setup is quite different to the present study.

The iterative method proposed in this work is applied to hybrid testing of a cantilever beam. This choice is inspired by the testing of aircraft wings, which, due to the limited capabilities of typical-sized wind tunnels, might not fit the experimental facility. Normally, at the root the behaviour of wings and fuselage can be reliably simulated with a numerical model of both structure and aerodynamics. However, the complex aerodynamic phenomena at the tip are much harder to simulate, and physical experiments are usually needed. Then one might want to replace the wing portion close to the root with a numerical model and perform a hybrid test on the wing portion containing the tip. 
Although attempts at hybrid testing of flexible wings can be found in the literature \cite{FagleyAero,SuSongAero}, in these experiments interface errors are not monitored, so they should not be regarded as a fully coupled RTHT experiment in the sense intended here.
The state of the art on HT is not mature enough to ensure the feasibility of such an ambitious test in which interface forces need to be matched up to a given accuracy without the assurance of interface measurements.
%, therefore the imposed forces need to be measured from a flexible system often having low damping. 

To assess the feasibility of hybrid testing of wings, we test a prototype case where the aerodynamics is neglected and the focus is on the structural dynamics. 
%Moreover, the torsional bending behaviour of wings would require three degrees of freedom at the interface whereas we want to restrict the test to two. 
We choose to target the bending motion of a cantilever beam with two degrees of freedom at the interface; torsional effects are ignored. While tests have been conducted on two or more interface degrees-of-freedom and on lightly damped structures (see, for instance, \cite{AbbiatiMdof,FermandoisMdof2017,NajafiMdof2020}), both of these combined with the direct force measurement challenge makes this a particularly difficult type of test.

\section{Theory and methods}
\label{sec:2}
In this section, the general framework for hybrid testing will be described, starting from the derivation of dynamic substructuring, moving to the description of RTHT and the concept of substructurability. Finally, the iterative method proposed in this work will be detailed.

\subsection{Dynamic substructuring}
\label{sec:ds}
Dynamic substructuring consists of identifying, from either first principles or physical experiments, all the components of a structure to then compute the dynamics of the assembled structure by coupling them together~\cite{Rixen}. In this sense, it differs from the idea of RTHT where no identification of the PS is sought but rather one wants to test the PS under real-time coupling conditions between PS and NS. However, some of the equations derived in the context of substructuring will be used in this work, and for this reason, a short description of them in the context of RTHT is provided in this section. The notation adopted here is the one proposed in~\cite{Allen}, where the interested reader can find more details on dynamic substructuring.
For illustration purposes, we assume that a model of the whole original structure is known, even though in reality the physical substructure is not identified, so there is no model available for it. For the sake of simplicity, we also assume the original structure to be linear and discretised with an appropriate discretisation, e.g. finite elements. As will be shown later, the proposed method presented in \autoref{sec:prop} can be easily applied to nonlinear structures but the same cannot be said about the general theory of dynamic substructuring nor RTHT. In fact, the literature on nonlinear dynamic substructuring is still scarce \cite{Allen}. Similarly, most of RTHT experiments are performed under the assumption of linearity or linearised around the operating conditions. For this reason, we restrict ourselves to the presentation of the linear case and then in \autoref{sec:prop} we will extend to the nonlinear one.

The equation of motion of the original structure can be written as a system of linear time invariant ODEs as
\begin{equation}
    M \ddot{u}(t)+C\dot{u}(t)+Ku(t)=f^{(e)}(t),
\end{equation}
where $u(t)$ is the state vector, $M$, $C$, $K$ are the mass, damping, and stiffness matrices of the structure and $f^{(e)}(t)$ a generic external forcing vector.

In the Laplace domain, we can define the dynamic stiffness matrix as
\begin{equation}
    D(s) = s^2M+sC+K,
\end{equation}
so that the original equation of motion can be written in the Laplace domain simply as
\begin{equation}
    D(s)U(s) = F^{(e)}(s).\label{eq:compact_orig}
\end{equation}

In dynamic substructuring, the original structure is divided into two or more substructures and the dynamics of each of them is studied separately and subsequently assembled \cite{Rixen}. In the context of hybrid testing, although the case of multiple substructures has been treated \cite{AbbiatiMdof}, the most common setup consists of only two substructures, the numerical one (NS) and the physical one (PS), which will be denoted in the following by the subscripts $N$ and $P$ respectively. The original structure is split into two at an interface that contains one or more degrees of freedom. Both in the case of an HT experiment and in the case of experimental identification of a substructure in dynamic substructuring, all the degrees of freedom of the interface must be actuated independently, so, typically, the selected interface contains a low number of degrees of freedom.
%The degrees of freedom of the original structure are then split into four vectors: the displacement in the bulk of the substructures, 
%The split is done on an interface between the two, so the degrees of freedom 
%The original degrees of freedom are then split into 

Within a substructure, one can then distinguish between degrees of freedom in the bulk of the substructure, which will be denoted by $b$, and those at the interface, which will be denoted by $i$. 

The equilibrium equations for the numerical substructure can then be written as
\begin{subequations}\begin{align}
    &D_{Nbb}(s) U_{Nb}(s)+
    D_{Nbi}(s) U_{Ni}(s) = F^{(e)}_{Nb}(s),\label{eq:Nb}\\
    &D_{Nib}(s) U_{Nb}(s)+
    D_{Nii}(s) U_{Ni}(s) = F^{(e)}_{Ni}(s)+F^{(i)}_{Ni}(s),\label{eq:Ni}
\end{align}\label{eq:eq}\end{subequations}
where \autoref{eq:Nb} and \autoref{eq:Ni} represent the equilibrium in the bulk and at the interface respectively. 
Similarly for the physical substructure:
\begin{subequations}\begin{align}
&D_{Pbb}(s) U_{Pb}(s)+
    D_{Pbi}(s) U_{Pi}(s) = F^{(e)}_{Pb}(s)\label{eq:Pb}\\
&D_{Pib}(s) U_{Pb}(s)+
    D_{Pii}(s) U_{Pi}(s) = F^{(e)}_{Pi}(s)+F^{(i)}_{Pi}(s).\label{eq:Pi}
\end{align}\end{subequations}

In addition to the external forces $F^{(e)}$, the internal forces exchanged between the two substructures must be included, which are represented by the $F^{(i)}$ terms on the right-hand sides of the equilibrium equations of the interface (\autoref{eq:Ni} and \autoref{eq:Pi}). To fully replicate the behaviour of the original structure, the so called \textit{equilibrium condition}~\cite{Allen} has to be fulfilled, which implies that said internal forces must sum to zero:
\begin{equation}
    F^{(i)}_{Ni}(s) + F^{(i)}_{Pi}(s) = 0.\label{eq:com1}
\end{equation}

At the same time, the degrees of freedom at the interface must satisfy the so-called \textit{compatibility condition}~\cite{Allen}
\begin{equation}
    U_{Ni}(s) = U_{Pi}(s)\label{eq:eq1}
\end{equation}

By enforcing both equilibrium and compatibility conditions, the behaviour of original system can be fully retrieved by assembling  \autoref{eq:Nb}, the sum of \autoref{eq:Ni} and \autoref{eq:Pi}, and \autoref{eq:Pb} into a single system:
\begin{equation}
\begin{bmatrix}
    D_{Nbb}(s) &D_{Nbi}(s) &       0       \\
    D_{Nib}(s) &D_{Nii}(s)+D_{Pii}(s) &D_{Pib}(s)\\
        0   &D_{Pbi}(s)         &D_{Pbb}(s)
\end{bmatrix}
\begin{bmatrix}
    U_{Nb}(s)\\U_{Ni}(s)\\U_{Pb}(s)
\end{bmatrix}
= 
\begin{bmatrix}
    F^{(e)}_{Nb}(s)\\F^{(e)}_{Ni}(s)+F^{(e)}_{Pi}(s)\\F^{(e)}_{Pb}(s).
\end{bmatrix}
\label{eq:orig}
\end{equation}
where the internal forces cancelled out thanks to the equilibrium condition and the interface displacements at the physical side have been substituted by those at the numerical side using the compatibility condition. This system is then equivalent to the original system~\autoref{eq:compact_orig}.

In a hybrid testing experiment, the degrees of freedom in the bulk of the PS are not accessible, since only forces and displacement at the interface are typically measured. Similarly, the NS only interacts with the PS through interface quantities. For what will be discussed later, it is useful to derive the relationship between interface forces and interface displacements for both substructures as they will represent their transfer functions. Such relationships can be found by expressing the bulk degrees of freedom from \autoref{eq:Nb} and \autoref{eq:Pb} as a function of the interface degrees of freedom and substituting them into \autoref{eq:Ni} and \autoref{eq:Pi}, which leads to
\begin{subequations}\begin{align}
&  D_N(s)
    U_{Ni}(s) = F^{(e)}_{N}(s)+F^{(i)}_{Ni}(s),\label{eq:eqN}
    \\
&  D_P(s)
    U_{Pi}(s) = F^{(e)}_{P}(s)+F^{(i)}_{Pi}(s),\label{eq:eqP}
\end{align}\end{subequations}
where the transfer function matrices $D_N(s)$ and $D_P(s)$ obtained via condensation on the interface degrees of freedom are
\begin{subequations}\begin{align}
&  D_N(s) = D_{Nii}(s)-D_{Nib}(s)    D_{Nbb}^{-1}(s)D_{Nbi}(s),
    \\
&  D_P(s) = D_{Pii}(s)-D_{Pib}(s)    D_{Pbb}^{-1}(s)D_{Pbi}(s),
\end{align}\label{eq:tf}\end{subequations}
and the condensed external force vectors $F^{(e)}_{N}(s)$ and $F^{(e)}_{P}(s)$ are
\begin{subequations}\begin{align}
&   F^{(e)}_{N}(s) = -D_{Nib}(s)D_{Nbb}^{-1}(s)F^{(e)}_{Nb}(s) +F^{(e)}_{Ni}(s),
    \\
&  F^{(e)}_{P}(s) = -D_{Pib}(s)D_{Pbb}^{-1}(s)F^{(e)}_{Pb}(s) +F^{(e)}_{Pi}(s).
\end{align}\label{eq:fe}\end{subequations}

It is possible to observe that, both in the definition of the condensed transfer functions of \autoref{eq:tf} and in that of the condensed external forces of \autoref{eq:fe}, the inversion of the matrices $D_{Nbb}$ and $D_{Pbb}$ appears. These matrices represent the dynamic stiffness matrices of the NS and PS when the degrees of freedom of the interface are clamped, which will be referred to as the \textit{clamped-interface} substructures in the remainder. The eigenvalues $\lambda_k$ of the \textit{clamped-interface} substructures are stored in the matrices $D_{Nbb}$ and $D_{Pbb}$ in the form of polynomials in $(s-\lambda_k)$. Due to the inversion operation, these eigenvalues become poles of the $D_N(s)$ and $D_P(s)$ transfer functions, meaning that for a small displacement at the interface, high interface forces can be generated if a resonance of the \textit{clamped-interface} substructure is excited. %If the real part of these eigenvalues is high enough, such resonances of the clamped substructures have moderate amplitude but if the damping in the system is low, they can compromise the results. 
It will be shown in the experimental results section that this can be indeed an issue during a hybrid test. In fact, these resonances can either generate a high-amplitude response in the PS, thus threatening the structural integrity of the test rig, or an ill-conditioned numerical problem. This consideration also emphasises the importance of the preliminary design of the test, where an excitation at the same frequency of these resonances could be avoided, for example, by changing the location of the interface.

In the remainder of this article, only the interface degrees of freedom and the interface forces will be considered. For readability, in the following the subscript $i$ on the variables $U_{Ni}$, $U_{Pi}$, $F_{Ni}^{(i)}$, and $F_{Pi}^{(i)}$, will be omitted. We now move on to the description of how the compatibility and equilibrium conditions are experimentally enforced in a real-time hybrid test.

\subsection{Generic scheme of a real-time Hybrid Test}
\label{sec:RTHT_generic}
In an RTHT, the two substructures, NS and PS, must be coupled together by physically controlling the behaviour of the interface. The aim of the test is to impose the compatibility and equilibrium conditions of \autoref{eq:com1} and \autoref{eq:eq1} in real time.
% \begin{subequations}\begin{align}
% &  U_{N}(s)=U_{P}(s),\label{eq:eq1}\\
% &  F^{(i)}_{N}(s)+F^{(i)}_{P}(s)=0.\label{eq:com1}
% \end{align}\end{subequations}

%Sensors placed at the interface of the PS allow to measure forces and displacements. The experiment can be either force controlled or displacement controlled. In fact, either the measured displacement $U_P(s)$ is imposed on the numerical structure thus automatically satisfying \autoref{eq:eq1}, then the resulting internal force is imposed on the PS, thus satisfying \autoref{eq:com1}.

For the sake of simplicity, we will treat the case where the PS is displacement controlled, but the force-controlled case can also be treated in the same fashion by simply switching force and displacement signals in the control scheme. For the sake of generality, we will refer to a generic test with multiple controlled degrees of freedom at the interface between PS and NS, although in practice the number of degrees of freedom at the interface is often just one.

\begin{figure}[ht]
    \centering
    \includegraphics[scale=0.27]{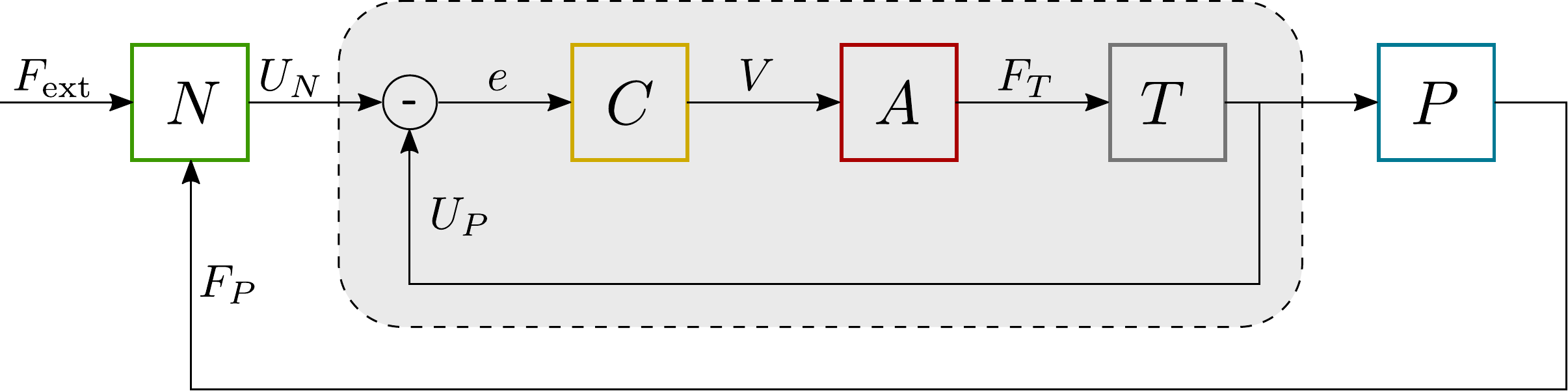}
    \caption{Generic scheme of a RTHT: $N$ numerical substructure, $C$ control algorithm, $A$ actuators, $T$ transfer system, $P$ physical substructure. The grey block represents the whole control system.}
    \label{fig:con_sch}
\end{figure}

In \autoref{fig:con_sch}, a generic setup of an RTHT experiment in displacement control is depicted. The transfer function $D_N$ of the NS is known, and so is the external force vector on the NS $F_{N}^{(i)}$. The displacement vector $U_P$ and the internal force vector $F_{P}^{(i)}$ are measured with sensors placed at the PS interface. If the interface is not directly accessible, sensors could be placed on the transfer system and the interface quantities reconstructed from those of the transfer system. The reconstruction of the interface displacement is typically straightforward, whereas that of the force requires an identification of the transfer system dynamics, as discussed in the experimental setup section. 

The interface force $F_P^{(i)}$ is measured from the sensors at the physical interface and, together with a chosen external forcing $F_{N}^{(e)}$, applied numerically to the NS. The imposition of the measured $F_P^{(i)}$ on the NS coincides with enforcing the equilibrium condition of \autoref{eq:eq1}. By solving the numerical model of the NS of \autoref{eq:eqN}, the commanded interface displacement $U_N$ is obtained. To satisfy the compatibility condition \autoref{eq:com1}, this displacement must match that measured at the PS, $U_P$, meaning that the command signal $U_N$ must be imposed on the PS in real-time. If the command signal $U_N$ is instantaneously imposed on the PS, then the equilibrium conditions are satisfied and the hybrid structure fully replicates the original. Combining the equations of motion of NS and PS condensed at the interface (\autoref{eq:eqN} and \autoref{eq:eqP}) with the compatibility and equilibrium equations (\autoref{eq:eq1} and \autoref{eq:com1}), the relationship between external forcing and displacement at the interface simply reads:
\begin{equation}
    \left(D_N(s)+D_P(s)\right)U_{N}(s)
    = F^{(e)}_{P}(s)+F^{(e)}_{N}(s),
    \label{eq:tf_hs}
\end{equation}
which is equivalent to the equation of motion of the original structure (\autoref{eq:orig}) condensed on the interface degrees of freedom.

The imposition of the command signal on the PS is done by the control system, which is generically composed of sensors, actuators, the transfer system, and the control algorithm. Most of the literature on RTHT is devoted to the design of sophisticated control systems that implement state-of-the-art control strategies to maximise the accuracy of the realisation of the command signal. As will be clarified in \autoref{sec:prop}, said control strategies are not required in the framework of the proposed method, so, for the sake of simplicity, here we report the architecture of the simplest possible control system, the feedback loop highlighted in grey in \autoref{fig:con_sch}. In this loop, the two signals $U_N$ and $U_P$ are compared and the discrepancy $e$ between the commanded and the actual signal is fed to a generic control algorithm (e.g., PID) $C$ that sends a voltage $V$ to the actuators $A$; the actuators impose a force $F_T$ on the transfer system $T$ that controls the PS to minimise the error $e$. The experiment then requires two nested loops, the inner one composed of the control algorithm, actuators, and the transfer system, and the outer one composed of NS, PS, and control system. 

Since the aim of the test is to realise the outer loop in a single time step, the inner loop has to realise the command signal in a much shorter time. As a matter of fact, an instantaneous realisation of the command signal is never possible, so the fastest time-scale that the hybrid test can accurately replicate is limited by the capabilities of the control system. In the design stage of a RTHT experiment, it is then crucial to assess the effect that an imperfect realisation of the command signal would have on the dynamics of the hybrid structure. To this end, in \cite{Terkovics2016}, the concept of substructurability has been introduced, which will be detailed in the next section.

\subsection{Substructurability}
\label{sec:RTHT_delay}
The substructurability of a RTHT experiment has been defined in \cite{Terkovics2016} as the system's tolerance to delays. Similarly, in \cite{Maghareh2017}, predictive indicators of the stability and performance of a RTHT experiment have been proposed to account for delays in the outer loop.
In fact, the control system introduces various distortions in the signal that, in addition to affecting the signal amplitude, are in the form of delays and lags. 

Several sources of delay are present in an RTHT experiment, such as the time to compute the response of the NS, the sampling time, and most importantly, the dynamics of actuators and transfer system and the delay introduced by the feedback loop. The first two are negligible compared to the latter, so it is common to assume the imposition of forces and the computation of the command signal to be instantaneous and to condense all sources of delays into a single one that retards the realisation of the command signal. 

\begin{figure}[ht]
    \centering
    \includegraphics[scale=0.27]{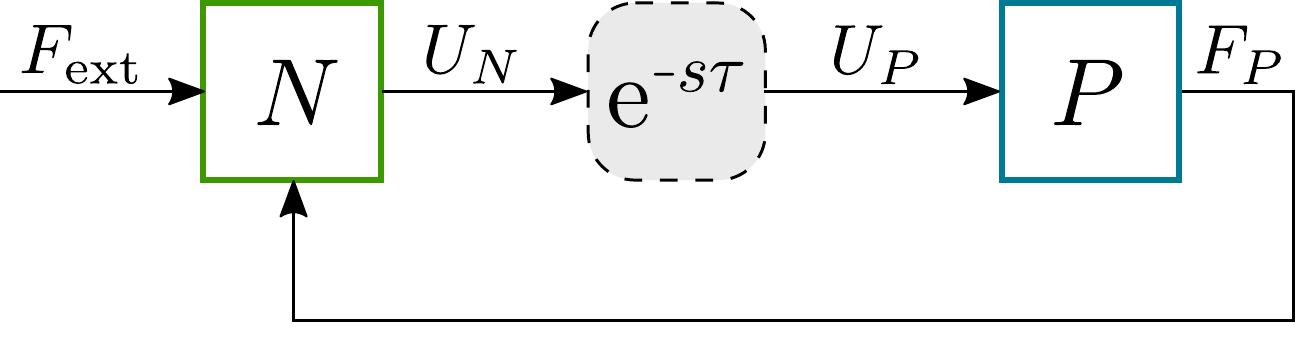}
    \caption{RTHT scheme under the pure delay assumption.}
    \label{fig:csd}
\end{figure}

In \cite{Wallace2005}, delay differential equation models have been proposed to assess the fidelity of an RTHT experiment where the inner feedback loop of the control system is replaced by a pure delay. In general, the feedback loop delay and the actuator lags are frequency dependent, but, in the moderate frequency range of a structural vibration test, it is reasonable to approximate them as a constant time delay. Moreover, the information that can be extrapolated from such analysis would still give a good indication of the feasibility of the experiment without any prior knowledge or identification of the control system. A depiction of a generic RTHT scheme under the pure delay assumption is shown in \autoref{fig:csd}.
%This is an approximation because the distortions introduced by the control plant are not necessarily frequency proportional as in the delay assumption. The delay assumption makes sense at low frequencies, where the feedback loop delay and the actuators lags can be lumped together and treated as a constant time delay, even though the reality is more complicated and such delay is normally frequency dependent. The simplest possible configuration is then to replace the control plant by a pure delay and study the stability of the system in such conditions. 

Since we are considering the general case of multiple degrees of freedom of the interface, to understand the potential impact of interface errors on the overall system we will also assume that the same delay $\tau$ applies to every signal in $U_N$. Under the above mentioned assumptions, the imperfect realisation of the command signal simply reads
\begin{equation}
    U_{P}(s) = U_{N}(s)e^{-s\tau},
\end{equation}
from which the overall dynamics of the hybrid structure of \autoref{fig:csd} can be obtained and reads
\begin{equation}
    \left(D_N(s)+D_P(s)e^{-s\tau}\right)U_{N}(s)
    = F^{(e)}_{P}(s)+F^{(e)}_{N}(s).
\end{equation}

Compared to the original structure described by \autoref{eq:orig}, here the transfer function of the PS is multiplied by a delay transfer function. The dynamics of the hybrid structure can then become unstable due to the presence of delays. The term in brackets represents the transfer function of the hybrid structure, and its determinant defines the characteristic equation
\begin{equation}
    |D_N(s)+D_P(s)e^{-s\tau}| = 0.\label{eq:delay}
\end{equation}

Due to the presence of the exponential term, the characteristic equation has an infinite number of roots in $s$, each representing an eigenvalue of the hybrid structure. Since the value of $\tau$ is not known a priori, it is common to study the variations of the hybrid structure's eigenvalues as $\tau$ varies. In particular, one is normally interested in computing the stability boundaries, that is, the values of the delay $\tau$ for which the eigenvalues of the system cross the imaginary axis. This is done by solving the characteristic equation in $\tau$ while requiring $s$ to be purely imaginary. Crossings occur at different values of $\tau$, but the smallest delay that causes instability is the most significant and is termed the predictive stability indicator (PSI) in \cite{Maghareh2017}. Since in an RTHT experiment one has some freedom to choose the exact location of the interface, in \cite{Terkovics2016} a thorough investigation of how the stability boundaries are affected by said location is proposed. Based on such considerations, one could then design the RTHT with the most convenient interface location and make the hybrid structure stable for larger delays. 

Together with the problem of stability, the accuracy of the RTHT experiment is also a crucial aspect because the hybrid structure should ultimately replicate the original one with the highest possible fidelity. So, even if the hybrid structure is found to be stable, delay compensation techniques would still be needed to minimise the effect of the control system in the loop. For this reason, a major area of research in the RTHT literature is concerned with the design of sophisticated control schemes able to compensate for the delays and lags in the experiment. However, these strategies are not only cumbersome to implement, but also require the identification of each component of the control system and thus depend on the particular rig setup at hand \cite{Ruffini}. Moreover, as discussed in \cite{Maghareh2017}, a controller optimised for a certain PS might no longer be optimal for a different one, even if the control system stays the same due to the so-called control structure interaction (CSI). Generally speaking, most control schemes rely on the identification of the control system and in some cases also of the PS. 

In this work, a different perspective on RTHT is proposed, which is fully rig-independent and does not require any prior identification of the system, which will be presented in the next section.

\subsection{Proposed Iterative method}
\label{sec:prop}
%The method proposed in this work aims at resolving this problems by breaking the inner loop and replacing the real-time controller by an iterative scheme that works on the harmonic coefficients of the variables of interest. 
%  Witteveen second paper:  While in the last publication the subsystems had to be tested one after the other within a loop run, they can be processed totally independently from each other in this generalized method
Since the goal of most engineering tests is to capture the response of the structure at resonance, one common experimental method consists of testing the response of the structure to a periodic forcing, sweeping the forcing frequency across the range of interest and covering the resonant response. 
Unless the structure is so nonlinear that it bifurcates to quasi-periodic or chaotic states, its response will also be periodic. Under the assumption of periodic forcing and periodic response, one can then work in the Fourier domain rather than in the time domain and realise the hybrid test by balancing the harmonic coefficients of the signals rather than their values in time. Within this framework, the high performance of the control system that is necessary in a classical RTHT test is no longer required because the real-time constraint is lifted; instead, the Fourier coefficients that are used to generate the demand signal can be updated asynchronously over time.

%To better understand how the proposed method works, refer to \autoref{fig:con_sch}. The RTHT scheme depicted in \autoref{fig:con_sch} still applies to the case of the present method, but now the variables have to be seen as the Fourier coefficients of their corresponding time signals, rather than their Laplace representation. All the equations presented in \autoref{sec:ds}, are still valid here, with the simple substitution $s=\text{i}\Omega$, with the excitation frequency $\Omega$. Moreover, the controller $C$ is no longer real-time, but it now represents an iterative root finding scheme that can take as much time as required to converge to a solution.

For the sake of simplicity, let us assume that the original structure is linear, although nonlinearities can also be treated. Then the relation between the forces and displacement on the numerical side will be fully defined by the dynamic stiffness matrix representing the force-displacement response of the numerical structure condensed at the degrees of freedom of the interface. Applying the internal and external forces on the NS, one obtains the Fourier coefficients of the command displacement, which read
\begin{equation}
    U_N(\text{i}\Omega) = D_N(\text{i}\Omega)^{-1} \left(F_N^{(e)}(\text{i}\Omega) - F_P^{(i)}(\text{i}\Omega)\right).
\end{equation}
This equation automatically fulfils the equilibrium equation so, to close the loop, the compatibility condition has also to be fulfilled, which read
\begin{equation}
    {U}_P(\text{i}\Omega) = D_N(\text{i}\Omega)^{-1} \left(F_N^{(e)}(\text{i}\Omega) - F_P^{(i)}(\text{i}\Omega)\right).
\end{equation}

In this equation, $D_N$ and $F_N^{(e)}$ are defined by the user, so they do not change for fixed $\Omega$. As per $U_P$ and $F_P^{(i)}$, they are computed using a Fourier transform of the time signals measured from the interface sensors. Provided that the external forcing on the PS $F_P^{(e)}$ is at steady state, the harmonic coefficients $U_P$ and $F_P^{(i)}$ only depend on the harmonic coefficients of the actuator voltage $V$.

The vector $V$ is then the free variable that may be used to drive the residual vector $R$ to zero, where
\begin{equation}
    R(\text{i}\Omega,V)= {U}_P(\text{i}\Omega,V) - D_N(\text{i}\Omega)^{-1} \left(F_N^{(e)}(\text{i}\Omega) - F_P^{(i)}(\text{i}\Omega,V)\right).\label{eq:res}
\end{equation}

Here, the implicit dependency on $V$ of the measured variables $U_P$ and $F_P^{(i)}$ is made explicit. This is a system with as many algebraic equations as the degrees of freedom of the interface times the number of harmonic coefficients, in the variable $V$. Such a system of equations can then be solved with root-finding methods such as Newton-Raphson. In particular, since the Jacobian of the measured variables with respect to the voltage is not available, a Jacobian-free method such as Broyden's Method~(\cite{Broyden1965}) could be used to construct the Jacobian matrix or its inverse, within the iteration.

{If the actuator system is nonlinear, with the PS and NS remaining linear, the method could be extended to include higher harmonics in the voltage in order to suppress the higher harmonics in the measured interface force or displacement. Whether or not higher harmonics should be included in the testing of a linear HS only depends on how nonlinear the actuators/transfer system is, which can be easily detected by checking how well the single-harmonic approximation fits the measured signals. The case of a nonlinear structure is slightly different because the residual of \autoref{eq:res} has to be written for each harmonic, taking into account cross-harmonic interactions generated by the nonlinearities. If only the PS is nonlinear, then the residual for each harmonic up to the highest selected harmonic $N_H$ simply reads:
\begin{equation}
    R(\text{i}k\Omega,V)= {U}_P(\text{i}k\Omega,V) - D_N(\text{i}k\Omega)^{-1} \left(F_N^{(e)}(\text{i}k\Omega) - F_P^{(i)}(\text{i}k\Omega,V)\right),\quad k\in(0,N_H)\label{eq:res_h}
\end{equation}
as the cross-harmonic terms generated by nonlinearities in the PS are automatically taken into account in the measured force signals. Conversely, if the NS is nonlinear then a multi-harmonic dynamic stiffness matrix of the NS that takes into account the NS nonlinearities has to be defined, e.g., using the classical harmonic balance method.}

{The only major limitation of the proposed method is that it can only be applied to testing the periodic steady state of a given structure. However, this is a quite common occurrence in mechanical engineering, where structures are usually tested around their resonances at steady state. Aside from this requirement, the advantages of the method compared to RTHT are multiple. Due to the periodic steady-state assumption, the requirement of the controller to work in real time is lifted without enlarging the time scale of the outer loop. Moreover, the implementation is rig and PS independent, i.e., it does not have to be tuned to the physical components of the experiment. In contrast, control schemes based on model predictive control and inverse modelling are rig-specific and might require cumbersome identification procedures. With the present methodology, the only identification required is in the case where force sensors cannot be placed directly at the interface. In such a situation, identification of the transfer system would be required in order for the interface forces to be reconstructed from sensed forces and displacements. }

\section{Cantilever beam in bending motion}
\label{sec:beam}
\begin{figure}[ht]
\subfloat[Original structure.]{
    \centering
    \includegraphics[scale=.27]{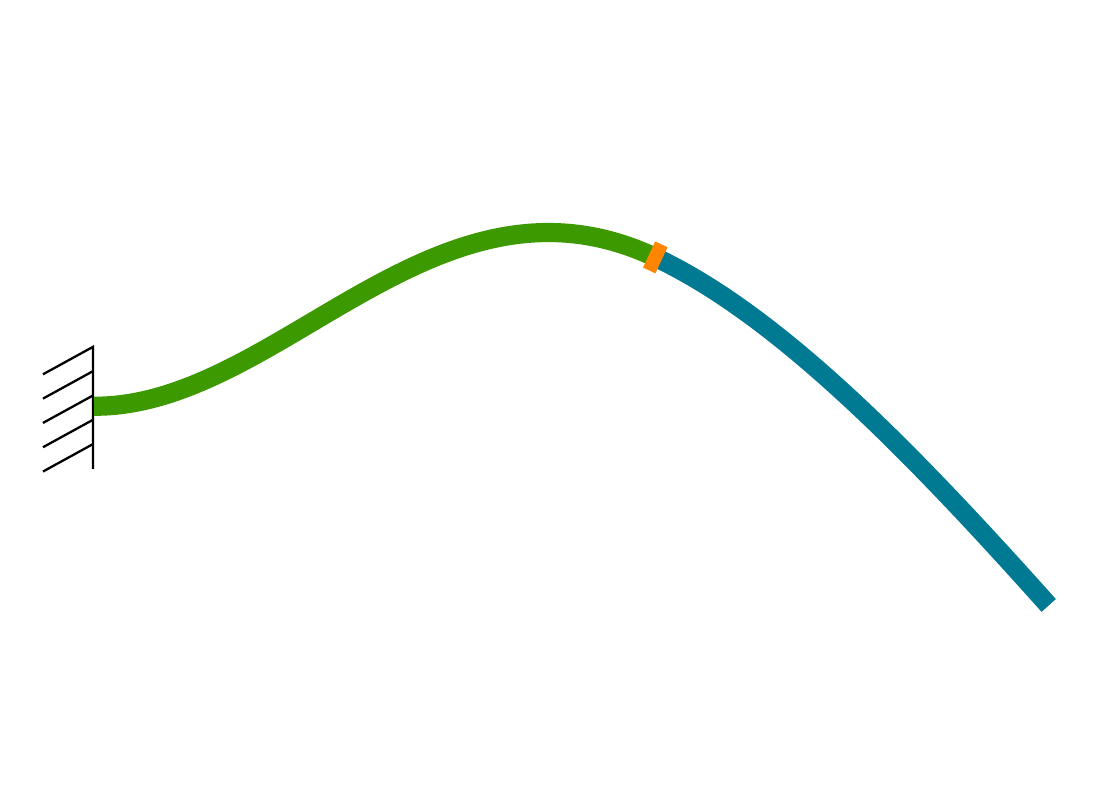}
}
\hfill
\subfloat[Hybrid structure.]{
    \centering
    \includegraphics[scale=.27]{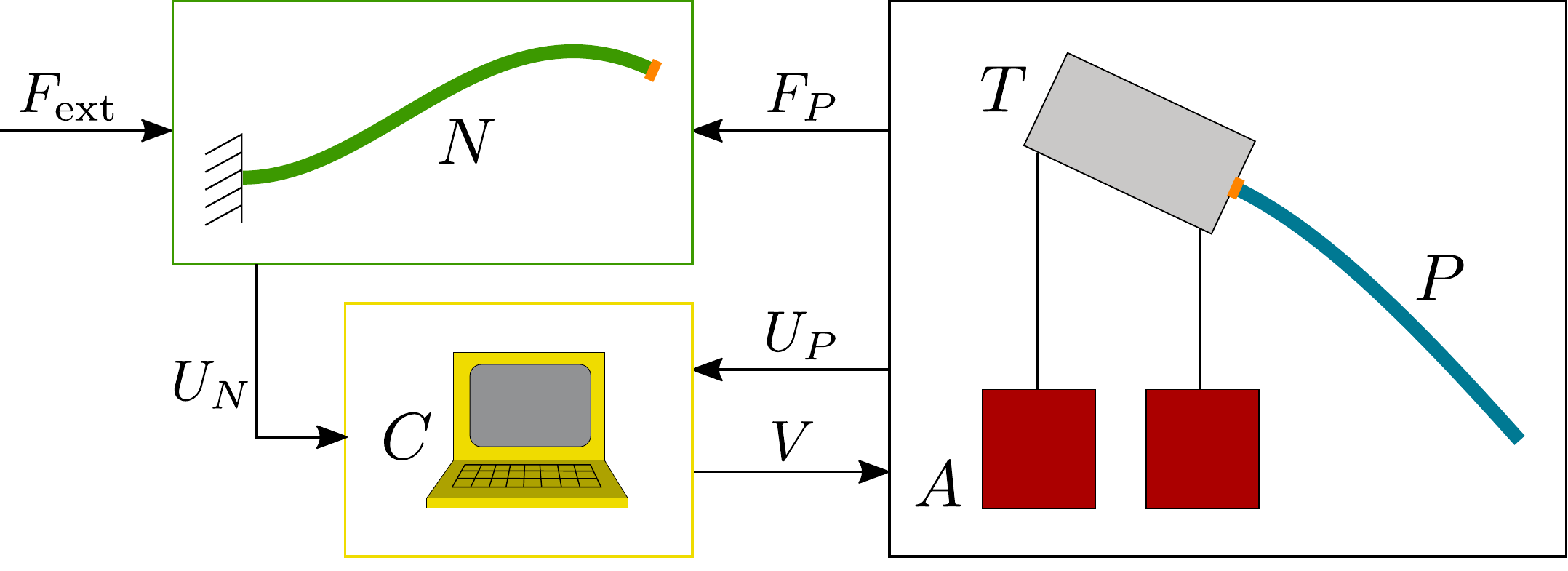}
}
\caption{Schematic of the original (a) and hybrid (b) cantilever beam in bending vibration. The numerical substructure ($N$) is denoted in green, the physical one ($P$) in cyan, and the interface in orange. The three blocks in the right plot represent the numerical structure, the displacement controller ($C$), and the test rig, which in turn comprises the actuators ($A$), the transfer system ($T$), and physical substructure.}
\label{fig:RTHT_cant}
\end{figure}

{The aim of this work is to perform real-time hybrid testing on a cantilever beam, with the numerical substructure being the clamped end and the physical substructure being the free end of the beam. We target a moderate vibration amplitude under which the bending motion of the cantilever beam can be considered linear. Since only the bending motion is of interest here, for any partition of the original structure, the interface only consists of two} degrees of freedom: the vertical displacement and the rotation of the section. A sketch of the original and hybrid structure is shown in \autoref{fig:RTHT_cant} where the split between NS and PS is shown together with the generic RTHT scheme applied in this example. Since there are two degrees of freedom at the interface, two actuators in the form of two shakers are required. This introduces an additional element of challenge in the experiment, as most of the hybrid testing setups deal with one controlled degree of freedom.

The hybrid structure represents the original $530$~mm long cantilever by splitting it at some point along its length into the NS (root end) and PS (tip end). Typically, in a hybrid test, the dynamics of the PS is unknown and no identification is required prior to the test. In this case, however, the hybrid structure is homogeneous, so, for the sake of elegance, we want the properties of the NS to roughly match those of the PS. Therefore, we ran a preliminary characterisation of the material properties of the PS to set the NS properties accordingly. Moreover, this preliminary identification provided us with an idea of what to expect in the hybrid test in terms of the natural frequencies and damping ratios of the hybrid structure. It is worth remarking that the dynamics identified during this preliminary step should not be seen as a reference solution for the hybrid test but more like a guideline for the choice of the NS properties and for the frequency range to sweep.% in that this preliminary identification is also flawed by factors such as an imperfect clamp. 

The physical structure is a steel ruler with a rectangular cross section of width $b=25.4$~mm and thickness $h=1$~mm. We choose to model its behaviour with linear beam theory and to treat dissipative phenomena in the simplest possible form, i.e., as a mass proportional and a stiffness proportional damping terms. This is a strong simplification because, from preliminary tests, we observed that the damping is not linear with the vibration amplitude. Again, since the focus here is not to have the NS perfectly matching the PS but to ensure the same range of behaviour; we favour simplicity of formulation over homogeneity. The equation of motion that describes the bending motion of the beam is
\begin{equation}
    E I u^{iv}(x,t) +\zeta_K E I \dot{u}^{iv}(x,t)+\zeta_M \rho A \dot{u}(x,t) +\rho A \ddot{u}(x,t) = 0
    \label{eq:pde}
\end{equation}
with clamped boundary conditions at $x=0$ and free boundary conditions at $x=L$, $L=530$~mm. The identified Young modulus is $E=217$~kg/ms$^2$/mm and the density is $\rho=8.21\cdot10^{-6}$~kg/mm$^3$. From the test, it is found that for the first modes of the beam, the damping ratios are much lower than $1\%$ at very low vibration amplitude and just below $1\%$ at vibration amplitude in the range that we plan to excite during the hybrid test. We then simply select $\zeta_M=0.0009$~1/ms, $\zeta_K=0.07$~ms to have a damping ratio in the first two modes equal to $0.8\%$ and higher on the other modes.

The first three modes of this cantilever beam have frequencies $2.8$~Hz, $17.6$~Hz, and $49.2$~Hz. Since the first mode is outside the range of the shakers used, we target the resonance of the second mode, as will be shown in the experimental results section. To cover this resonance, the operational frequency range of the test will be between $16$~Hz and $19$~Hz.

Having defined the NS properties and the operational frequency range, we now have to choose the location of the interface along the length of the beam. In classical real-time hybrid testing, the computation of stability boundaries as a function of interface location and interface delay gives an approximate indication of the substructurability of a given structure. The substructurability of the cantilever beam model is studied in the following section to highlight the challenges of the present testcase. With the proposed iterative method, the real-time constraint is lifted and so is the problem of introducing delays during the test. Nevertheless, some constraints on the choice of the location of the interface apply regardless of the chosen method. First, if one targets the resonance of a given mode, the displacement of the hybrid structure will be mainly along that mode shape. If there are nodes in the mode shape, placing the interface too close to a node would restrict the controllability of the test. In the present case, the targeted mode has only one node that is approximately $120$~mm from the free end, so a partition of the HS where the PS is around $120$~mm long should be avoided as the interface displacement would be very small and would therefore be overly affected by frictional effects.

When choosing the partition, another point to take into account is that one must avoid a match between the excitation frequency and a natural frequency of the \textit{clamped-interface} substructures, i.e. the natural frequencies of the NS and PS when the degrees of freedom of the interface are clamped. As discussed in \autoref{sec:ds}, the eigenvalues of the \textit{clamped-interface} substructures become poles of the dynamic stiffness matrices condensed at the interface; therefore, if the excitation frequency is tuned with one of them, a small interface displacement could generate high interface forces. Another way of seeing this is by recalling \autoref{eq:tf_hs}, which represents the relationship between interface displacement and interface forces in a perfect hybrid test. The resonant frequencies of the \textit{clamped-interface} substructures correspond to anti-resonances of the transfer function of the hybrid structure condensed at the interface, so the lower the damping, the more the interface will act like a node at that particular frequency, thus hindering the controllability of the experiment. Since we target the second mode of the HS, only the lower \textit{clamped-interface} modes of the two substructures can have a resonance in the operational range chosen, whereas if higher modes of the HS were of interest, the number of potential crossings would have been higher. In the present case, a PS long between $210$~mm and $230$~mm would have its first \textit{clamped-interface} mode between $16$~Hz and $19$~Hz, so a PS length in this range should also be avoided.

To be far from the node of the mode shape and from possible anti-resonances, the length of the PS in the experiment is chosen to be $L_P=360$~mm. With this choice, the first \textit{clamped-interface} modes of the PS will be at $6$~Hz, $40$~Hz, and $112$~Hz,  whereas all the \textit{clamped-interface} modes of the NS will be higher than $100$~Hz. Even with the precaution of avoiding that \textit{clamped-interface} modes of the PS match the excitation frequency, these modes could still potentially impact the test if tuned with multiples of the excitation frequency. In fact, the PS might receive a small component of higher harmonics of the excitation frequency if weak nonlinearities are present in the rig. In the experimental results section, it will be shown that a small second harmonic component in the force is observed in the time signal as a result of the excitation of the $40$~Hz \textit{clamped-interface} mode of the PS. 
%\footnote{{Doubt: including higher harmonics in the test should solve this problem but wouldn't their controllability be still limited because of the anti-resonance problem stated above?}}
%\footnote{{Other doubt: are the node and the anti-resonance issues pathology of the method or would they also appear in a classical RTHT? Technically they are associated with the problem of periodic excitation and they should not affect a random excitation but do they actually?}}

\subsection{Stability of real-time hybrid testing}
\label{sec:substruct}
Before moving to the experimental results obtained with the proposed method, we want to investigate the feasibility of classical RTHT on the present cantilever beam in the presence of small delays. The proposed method is not affected by delays in the control system, whereas in classical RTHT even small delays can potentially cause instabilities. Assessing the stability of a hybrid structure as a function of interface delays and interface location, i.e., its substructurability, can then provide a good indication of the feasibility of classical RTHT. Moreover, the largest admissible delay for which the hybrid structure is still stable provides a quantification of how challenging the test at hand is.
% how the interface position affects the test stability. 

The structure under study is the cantilever beam presented in the previous section, with parameters identified from the actual PS. To investigate the substructurability of this structure we discretise the PDE \autoref{eq:pde} with one-dimensional beam elements. Along the discretised beam, we choose the node corresponding to the interface and split the dynamic stiffness matrix of the original structure into two matrices for NS and PS. We then condense the dynamic stiffness matrices of the two substructures to obtain $D_N(s)$ and $D_P(s)$, as described in \autoref{sec:ds}. Defining the ratio between numerical and physical length $\alpha = L_N/L_P$, we want to compute the condensed matrices $D_N(s,\alpha)$ and $D_P(s,\alpha)$ as functions of $\alpha$. To do so, it is sufficient to have a fine enough discretisation and to choose a different node along the discretised beam so as to cover all possible discrete values of $\alpha$ between zero, a purely numerical beam, and one, a purely physical one. We also assume a displacement control experiment, with a constant delay $\tau$ equal for both actuators. Since the values of the delay are in the order of the millisecond (ms), in this section we work with this unit, therefore we express the frequency in kHz. Without going into details, suffice to say that each entry in the 2 by 2 matrices $D_N(s,\alpha)$ and $D_P(s,\alpha)$ is a rational function with polynomials in $s$ as the numerator and denominator, whose coefficients depend on $\alpha$. It follows that the determinant in \autoref{eq:delay} is also a rational function whose zeros coincide with the zeros of the numerator. The characteristic equation of the hybrid cantilever beam can then be rewritten as
\begin{equation}
    \mathcal{C}(s,\alpha,\tau) = \mathcal{P}_2(s,\alpha) e^{-2s\tau} + \mathcal{P}_1(s,\alpha) e^{-s \tau} + \mathcal{P}_0(s,\alpha)\label{eq:char} = 0,
\end{equation}
where each $\mathcal{P}_i(s,\alpha)$ denotes a generic polynomial in $s$ whose degree depends on the number of finite elements used in the discretisation. 

\autoref{eq:char} is a complex-valued scalar equation with two parameters, the delay $\tau$ and the partition ratio $\alpha$, and one complex-valued unknown $s$. The roots of~\autoref{eq:char} correspond to the eigenvalues of the hybrid structure and can be expressed as $s = \delta + 2 \pi \mathbf{i} f$. The imaginary part of each eigenvalue characterises the frequency $f$ of oscillation, and the real part $\delta$ determines the growth or decay of the oscillation, i.e., the stability. As discussed in \autoref{sec:RTHT_delay}, the stability of the hybrid structure can be studied by computing all the roots of the characteristic equation given $\tau$ and $\alpha$, and assessing whether any of them lies in the right-hand-side of the complex plane $(\delta-f)$, i.e., has positive-valued real part $\delta$. Due to the transcendental nature of \autoref{eq:char}, for each set of parameters $\tau$ and $\alpha$, infinitely many roots exist. However, rather than their specific values, one is typically interested in just knowing how many of them are unstable. Moreover, since the delay assumption does not hold for very high frequencies and since the actuators are limited in their frequency range, here we are only interested in knowing how many unstable roots there are that have an imaginary value that results in dynamics at a lower frequency than the cut-off for the shakers. In the present experiment, the admissible frequency range of the shakers is particularly high as it reaches up to $13$~kHz but in a typical rig the range can be much lower; we then choose a cut-off frequency of $0.5$~kHz.

Before delving into the solution of \autoref{eq:char}, it is worth making a few observations about its structure. First, notice that not only the delay term $e^{-s \tau}$ appears, but also its square, $e^{-2s \tau}$, feature that stems from to the presence of two degrees of freedom at the interface. Moreover, it is worth mentioning that both delay terms are multiplied by polynomials in $s$, which is the hallmark of a neutral delay differential equation (NDDE).
In fact, delays in the original delay differential equation affect not only the displacement but also its time derivatives. This is a feature of continuous structures because the split affects not only the stiffness term but also the inertia and damping terms. In terms of the arrangements of eigenvalues, the main difference between a delay differential equation (DDE) and a neutral one (NDDE) is the following: in the case of a DDE, it is possible to demonstrate that, from a certain frequency onwards, all the high-frequency roots are stable; in contrast, in the case of an NDDE, the high frequency roots accumulate on vertical lines of the complex plane~\cite{DDEbook}, i.e., constant $\delta$ lines. This means that if one of the accumulation lines lies on the right side of the complex plane, the system will have infinitely many unstable roots. In the present case, there will be two accumulation lines due to the quadratic nature of the characteristic equation in the delay terms. 

\begin{figure}[ht]
    \centering
    \includegraphics[scale=0.6]{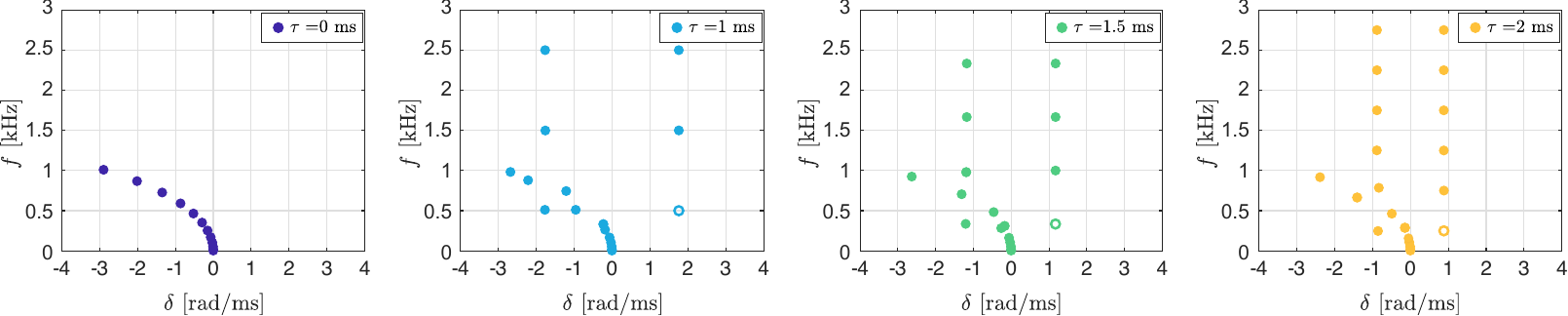}
    \caption{Eigenvalues of the characteristic equation of the hybrid cantilever with interface located in the middle of the beam ($\alpha = 0.5$) computed for four different values of the delay.}
    \label{fig:roots}
\end{figure}

Several techniques can be used to solve \autoref{eq:char}. Here, we use the bisection method and the MDBM Matlab code developed by Bachrathy and Stepan~\cite{MDBM}. In MDBM, a domain of existence for each parameter must be provided and the solution is sought within the specified range only. As a first step, we compute the roots of \autoref{eq:char} for four different values of $\tau = \lbrace 0,1.2,1.75,2.3\rbrace$ms, fixed $\alpha=0.5$, and for range values of $f \in [0,3]$kHz and $\delta \in [-4,4]$rad/ms. They are shown in \autoref{fig:roots}. The zero-delay system, coinciding with the original structure, has only stable roots, as expected. The organisation of these roots in the complex plane is dictated by the damping model used, which is, in this case, dominated by the stiffness proportional term. For nonzero values of the delay parameter, the roots appearing can be categorised into two families: those lying in a neighbourhood of the eigenvalues of the zero delay system, which seem to be departing from their zero delay counterpart as the delay increases; and a new set of roots with no counterpart in the zero delay system. None of the roots belonging to the first family becomes unstable in the selected range of parameter values tested in \autoref{fig:roots}, even though one can expect that for larger values of delay. Regarding the roots belonging to the second family, they lie in the neighbourhoods of two vertical lines, which suggests these lines might be where high-frequency roots accumulate. For any nonzero value of the delay, one of the two accumulation lines lies on the right side of the complex plane, meaning that the hybrid structure is always unstable with an infinite number of unstable roots. However, as already discussed, the behaviour of high-frequency roots is not the focus here as we are only interested in roots with a frequency lower than a selected cut-off. Focusing now on the unstable root with the lowest frequency, denoted by the empty dot in each plot of~\autoref{fig:roots}, notice that both its real and imaginary parts decrease with increasing delay.

\begin{figure}[ht]
    \centering
    \subfloat[Frequency-delay view.\label{fig:root_locus}]{
    \centering
    \includegraphics[scale=.6]{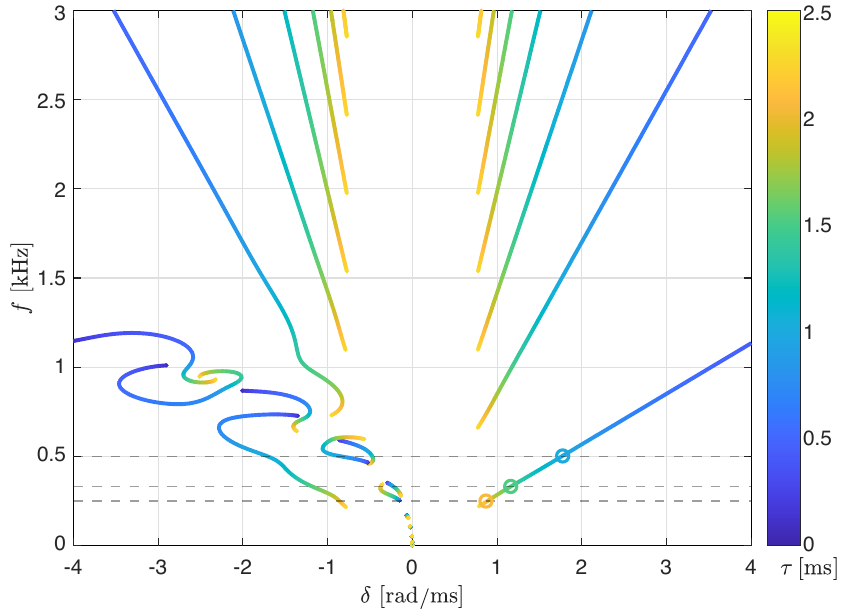}
    }       
    %\hfill
    \subfloat[Interface location-delay view.\label{fig:tau_alpha}]{
    \centering
    \includegraphics[scale=.6]{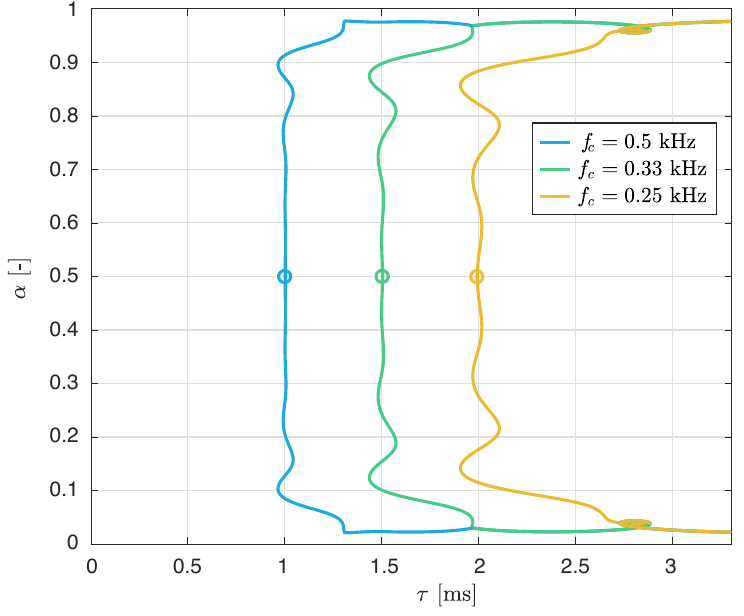}
    }
    \caption{Root locus of unstable eigenvalues of a hybrid cantilever beam for varying delay $\tau$, ratio of physical length over numerical length $\alpha$, and frequency $f$ with a cut off frequency of $500$~Hz. Four different values of the real part $\delta$~[1/ms] are reported in red scale. In the left plot the $\alpha-\tau$ view and in the right plot the $f-\tau$ view are given.}
\end{figure}

{To better understand how the roots behave for $\alpha=0.5$ still fixed but varying $\tau$ continuously, we now let $\tau$ be a free parameter and compute the solution of \autoref{eq:char} with $\tau \in [0,2.5]$ms. The root locus obtained is reported in \autoref{fig:root_locus}. Again, it is possible to distinguish the two families of roots mentioned above: the bottom left corner of~\autoref{fig:root_locus} contains the roots that originate from the eigenvalues of the zero delay system, whereas the straight lines in the plot correspond to the roots that do not have counterparts in the original system. The three empty dots of~\autoref{fig:roots}, representing the unstable roots with lowest frequency for $\tau=\lbrace 1,1.5,2\rbrace$ms respectively, are shown also in~\autoref{fig:root_locus}. They lie on the straight line corresponding to the locus of unstable roots with lowest frequency. Focusing on this straight line, it is possible to see that as $\tau$ decreases both the frequency and the real part of the root increase, following an inverse proportional behaviour with $f\approx 1/(2\tau)$. This means that, for any value of the delay, the system is always unstable. However, this instability occurs at increasingly high frequency; therefore it might not affect a real experiment if a cut-off is imposed. For instance, looking at the grey dashed lines in~\autoref{fig:root_locus}, placed respectively at $f=\lbrace 0.5, 0.33, 0.25 \rbrace$kHz, the critical values of delay at which the cut-off system becomes unstable are respectively $\tau = \lbrace 1, 1.5, 2\rbrace$ms. Since a selected cut-off frequency can always be imposed in a real experiment, it is interesting to see how the critical value of $\tau$ behaves as the interface location $\alpha$ varies. This will define a stability boundary for the cut-off system in the $(\tau-\alpha)$ space dividing the stable region of the parameter space from the unstable one. Fixing the frequency of the root to be equal to the selected cut-off, and letting $\tau$, $\alpha$, $\delta$ be free parameters, we can compute the stability boundary of the cut-off system. Such results are illustrated in~\autoref{fig:tau_alpha} where the three eigenvalues obtained for $\alpha=0.5$ are again reported as empty dots. Each line divides the $(\tau-\alpha)$ space into two regions, the one on the left of the line being stable. It is possible to notice that the interface location does not affect much the stability of the system for the most part, as the critical value of delay stays almost constant. Only in very small regions around $\alpha=0$ and $\alpha=1$ does the behaviour change and the cutoff system is stable for any delay. However, these regions are of little interest for a hybrid experiment, as they represent the trivial cases of a fully numerical or a fully physical structure. In general, the rule of thumb given by $f_{c} \approx 1/(2\tau_{c})$ holds for almost all interface locations. Therefore, if one has a rough estimation of the delay introduced by the control system, this rule provides an indication of the appropriate cut-off frequency to choose not to incur in an instability. }

The numerical findings presented in this section show that the current cantilever beam is particularly challenging to test in a classical RTHT framework. They also show the importance of such preliminary investigations before attempting a hybrid experiment. In fact, the restriction on the largest admissible delay imposes stringent requirements on the quality of the controller. Moreover, here we only treated the stability of the hybrid structure, but not its accuracy, compared to the original structure. For lightly damped structures such as the present one, it is expected that a small interface delay would mostly affect the overall damping in the system, thus potentially generating large errors in a resonant response. In fact, the assurance of stability should not be mistaken for the assurance of accuracy. 
Unlike classical RTHT, the proposed method is not affected by interface delays due to the iterative procedure. Moreover, the accuracy of the hybrid structure is only limited by sensing errors, as discussed in the next section.

% The collection of leftmost lines in the tau-alpha plane is so called stability boundary.
% there is always a pair of complex conjugate eigenvalue 
% Not to mention the positive tau case where everything is unstable.

\section{Experimental setup}
\label{sec:setup}
\begin{figure}[ht]
\centering
\subfloat[Lateral view of the rig.\label{fig:ska}]{\includegraphics[height=0.52\textwidth]{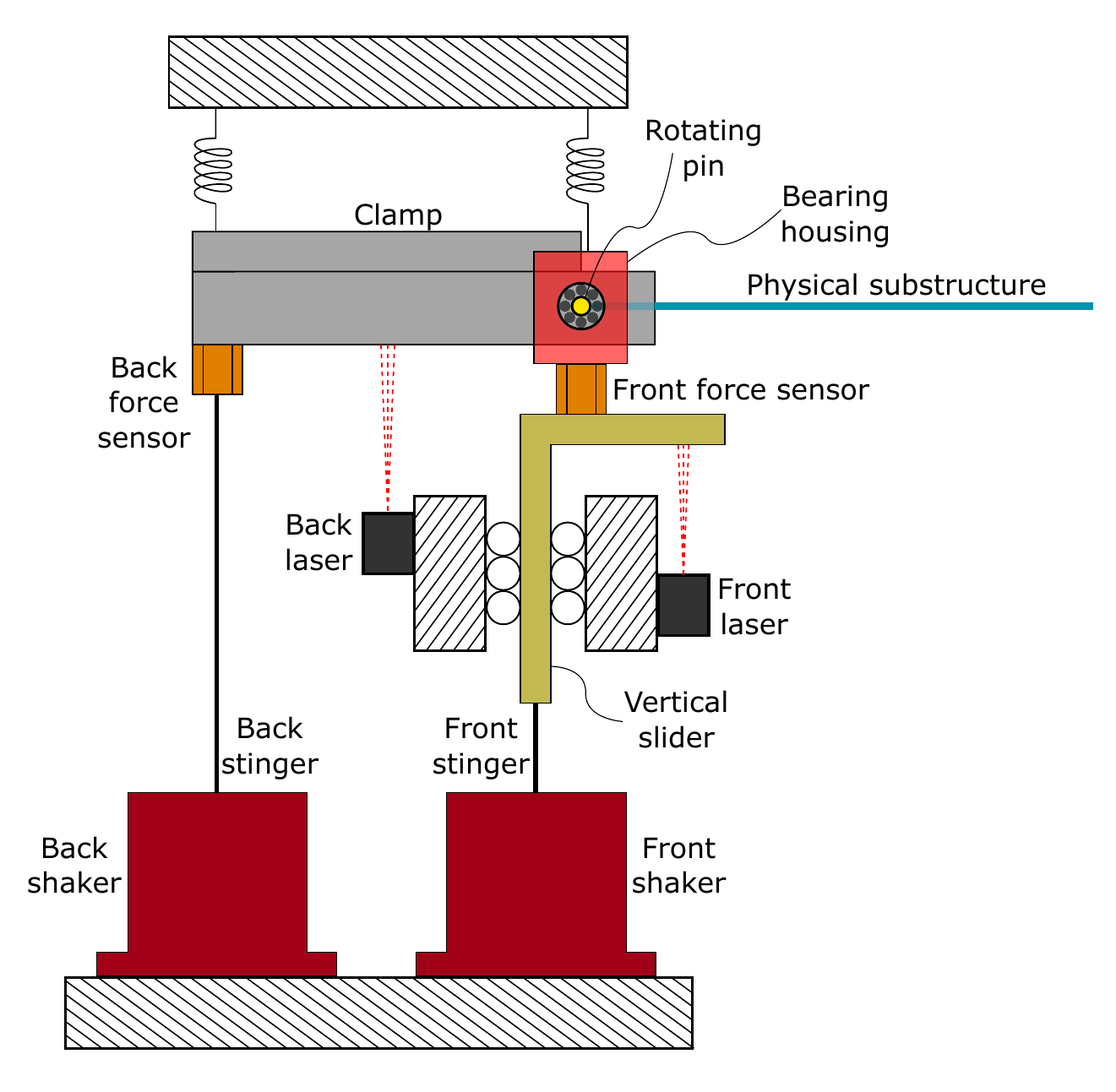}}
\hfill
\subfloat[Front view of the interface.\label{fig:skb}]{\includegraphics[height=0.52\textwidth]{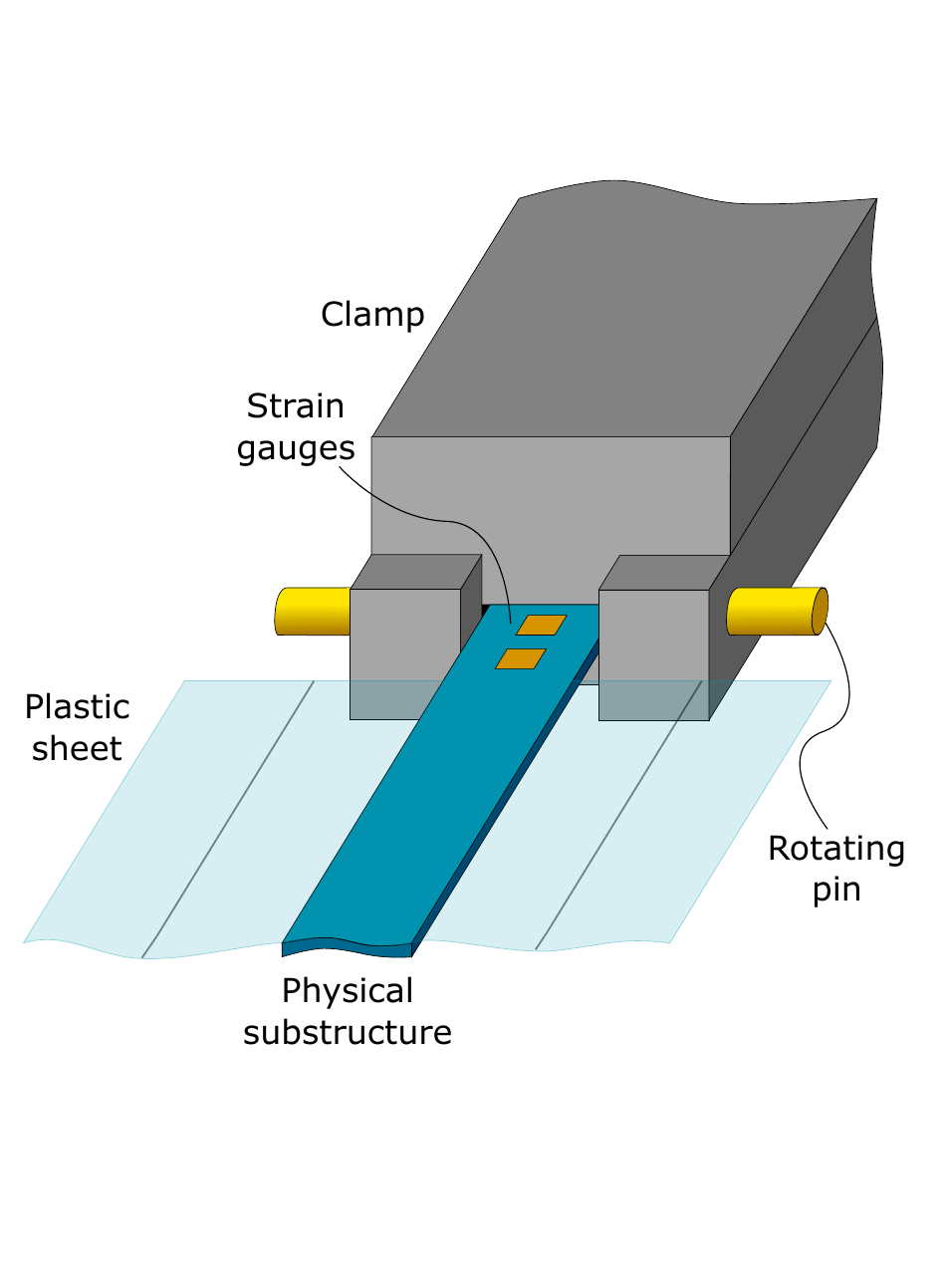}}
\caption{Schematic of the test rig from the side (a) and from the front (b) showing the transfer system holding the physical substructure, the actuation system composed by the two shakers, and the positions of sensors.} \label{fig:sk}
\end{figure}

\begin{figure}[ht]
\centering
\subfloat[Lateral view.\label{fig:cl1}]{\includegraphics[height=0.25\textwidth]{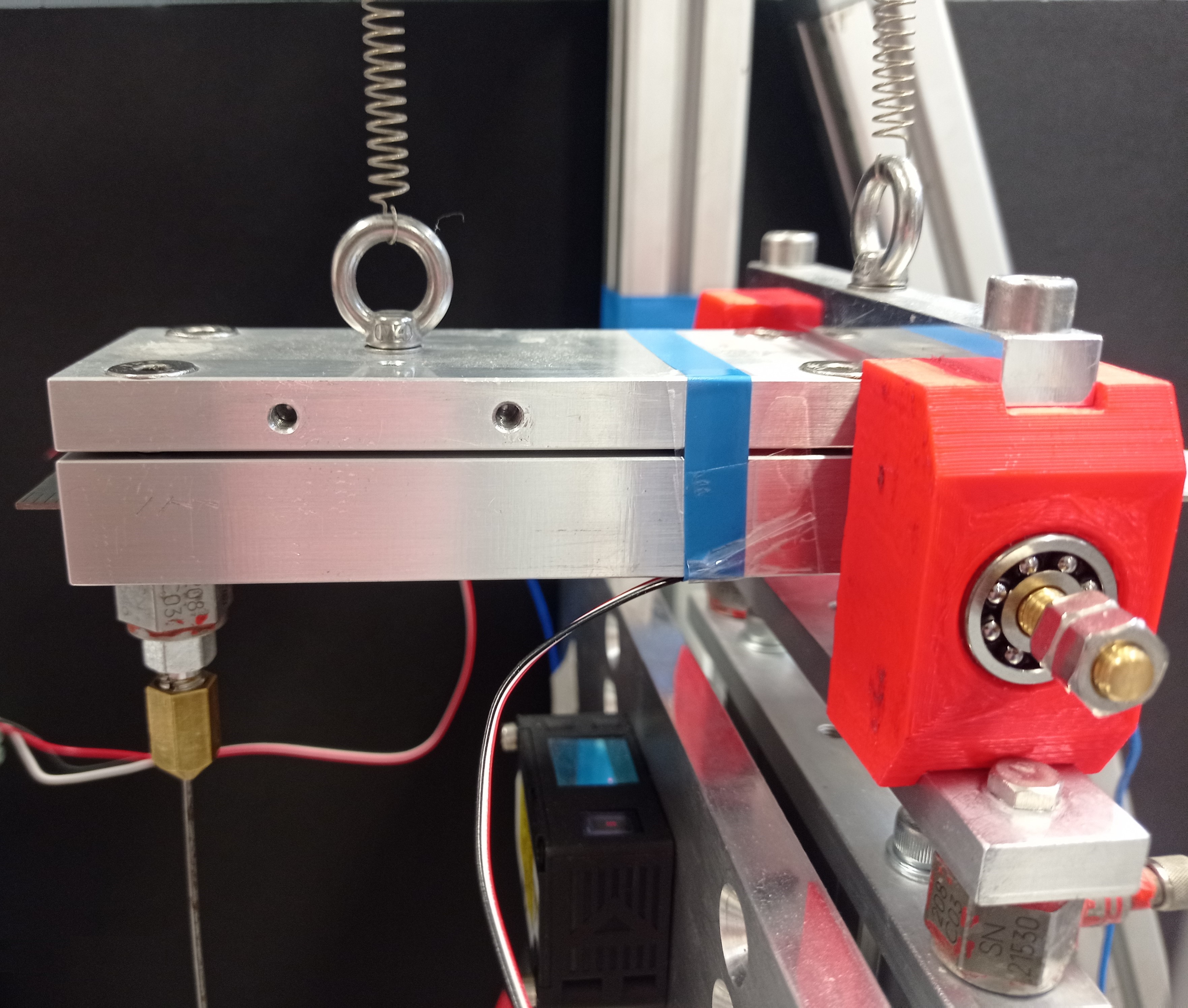}}
\hfill
\subfloat[Front view.\label{fig:cl2}]{\includegraphics[height=0.25\textwidth]{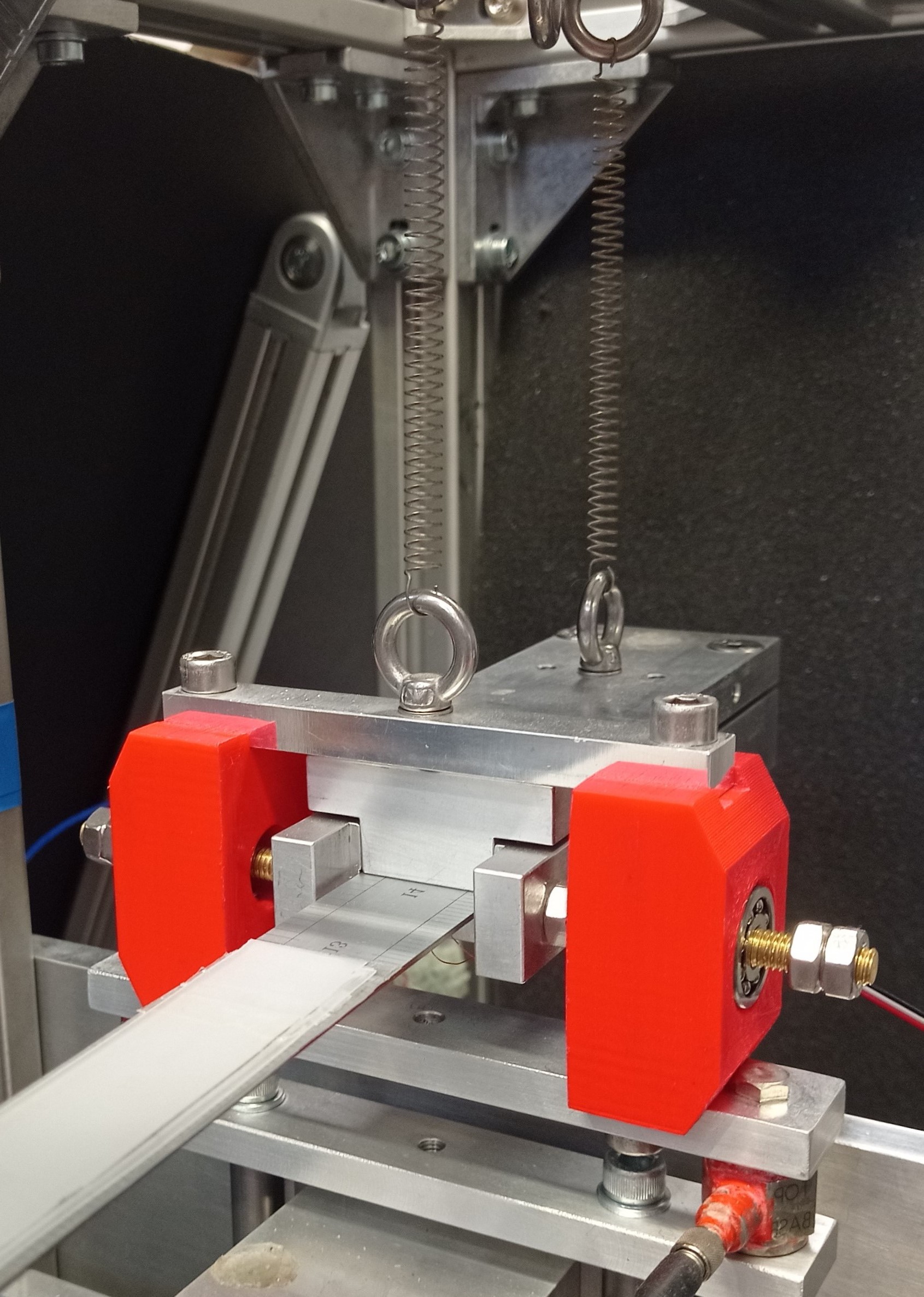}}
\hfill
\subfloat[Top view.\label{fig:cl3}]{\includegraphics[height=0.25\textwidth]{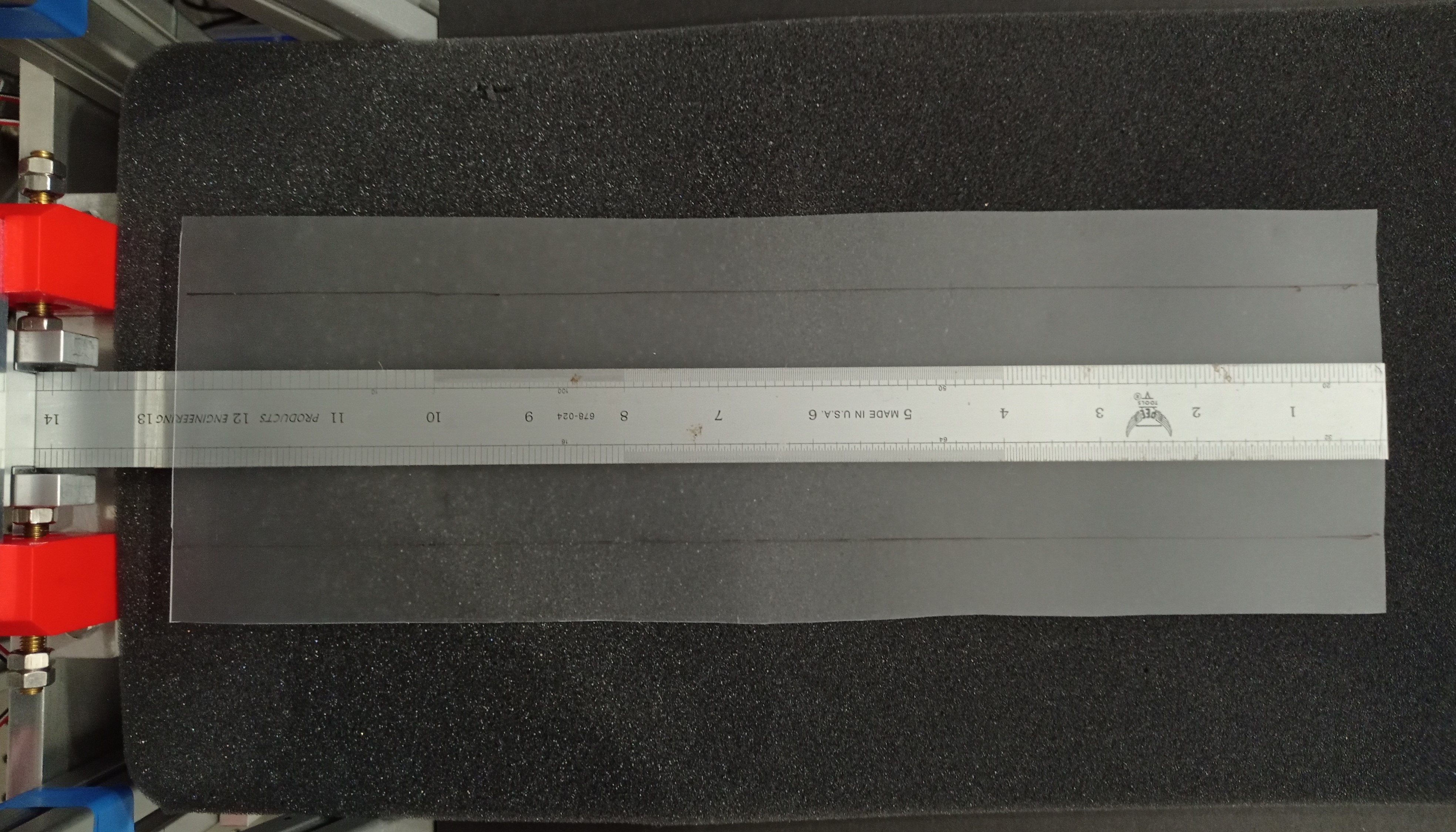}}
\caption{Lateral view of the clamp holding the cantilever beam, front view of clamp and interface, top view of the physical structure. The rigid plastic sheet attached to the physical substructure is shown in its narrower configuration in plot (b) and in its wider one in plot (c).} \label{fig:cl}
\end{figure}

We now provide a description of the experimental setup designed to perform the hybrid test. A depiction of the test rig is reported in \autoref{fig:sk}, and two views of the rig are shown in \autoref{fig:cl}. A $500$~mm long steel beam is mounted in a $140$~mm clamp and the remaining $360$~mm long portion of the beam corresponds to the physical substructure (PS). The clamp is supported by two springs from the top, and it is rigid enough to ensure that the portion of the beam enclosed in the clamp does not contribute to the flexibility of the PS. The section of the beam at the interface between PS and the clamp corresponds to the interface of the HS, whose motion can then be controlled by controlling that of the clamp. Two degrees of freedom, the vertical sliding and the rotation around the interface middle axis, are actuated by means of two LDS-50 shakers placed at the front and back of the rig.
%As mentioned, two actuators are necessary to control the two degrees of freedom of the interface, vertical sliding and . These are two LDS-50 shakers placed at the front and back of the clamp. 
To block the other four degrees of freedom, the front shaker is not directly attached to the clamp but to a vertical slider and a bearing housing rigidly mounted together. %and it sustains the purely vertical motion of the pin housing. 
The rotating pin inside the bearing is rigidly connected to the clamp and is perfectly coaxial with the middle axis of the interface, so that the latter can only rotate and translate vertically. Since the vertical slider is connected to the front shaker, the latter is entirely devoted to the control of the vertical motion of the interface. The back shaker is directly attached to the back end of the clamp, thus controlling the rotation. The stinger of this shaker is intentionally long to allow for moderate rotations without imposing significant side forces on the shaker or the clamp. %

As the physical structure is a simple steel beam, very low damping values were observed during the preliminary identification. To increase the robustness and to test the proposed method under varying damping conditions, the beam is complemented with an additional plastic sheet $127$~mm wide, glued on top of the beam. In this way, the aerodynamic damping of the PS increases without significantly affecting its elastic properties. By gradually reducing the width of the plastic sheet, different damping levels can be tested before testing the original bare beam configuration. Three different settings are planned, the most damped one with a $127$~mm wide layer of plastic as shown in \autoref{fig:cl}, a less damped one obtained by cutting along the black lines, thus removing two stripes of material from each side, and the least damped one obtained by cutting along the width of the beam. Each stripe is 25.4 mm wide, and it is glued back on top of the beam once cut in order not to alter the overall inertia. It is estimated that the widest and narrow plastic layers are able to respectively triple and double the damping as compared to the bare beam.
% configurations have approximately trice and twice as much damping as the bare beam respectively.

With respect to sensors, two lasers are used in the experiment; the first one, $l_1$, measures the position of the vertical slider, the second one. $l_2$, the vertical displacement of a point of the clamp 30 mm away from the pin axis. Defining $d_l=30$~mm as the distance between the two lasers, the interface displacement $u(t)$ and rotation  $\phi(t)$ can be reconstructed from the laser signals as
{\begin{subequations}\begin{align}
    & u(t) = l_1(t),
    \\
    & \phi(t) = \dfrac{l_1(t)-l_2(t)}{d_l}.
\end{align}\label{eq:reconstr_U}\end{subequations}

Two force sensors are placed, respectively, between the back stinger and the clamp, and between the vertical slider and the bearing housing. These could be used to reconstruct forces at the interface by subtracting the dynamics of all the components in between, i.e., the springs, the clamp, and the bearing. Such a choice would require an identification of this transfer system, which, in this case, is particularly challenging mainly due to the presence of the bearing, where friction effects are to be observed. A proper nonlinear identification of the transfer system would then be required, which is intrinsically challenging. Moreover, errors in the identified properties of the transfer system would propagate into the resulting properties of the hybrid structure. For small errors, it is expected that inaccuracies in the identification of the damping of the transfer system would mostly affect the damping of the hybrid structure, whereas those on mass and stiffness would affect the hybrid structure's natural frequencies. Intuitively, whenever the damping of the HS is very low as compared to that of the transfer system, its sensitivity to a small percentile error in the identification is very high. In the structure treated here this is particularly true because the HS is a simple steel beam with only material damping and little aerodynamic damping, whereas the transfer system contains a bearing which inevitably introduces frictional dissipation. In such situations, it is necessary to measure forces as close as possible to the interface. To do so, we propose to extract the interface shear force from strain measurements, using strain gauges placed directly on the beam, as detailed in the next section.

\subsection{Shear force measurement}
\label{sec:gauges}
In order to extract the shear force and moment at the interface, two strain gauges are placed on the PS in the immediate vicinity of the clamp. The bending moment is reconstructed by extrapolation from the gauges, whereas the shear force is reconstructed by finite difference in space. In particular, denoting with $d_1$ the distance from the interface of the closer strain gauge and $d_2$ the distance of the farthest one, the shear force $T(t)$ and the bending moment $M(t)$ are computed as
\begin{subequations}\begin{align}
    & T(t) = c \dfrac{\epsilon_2(t)-\epsilon_1(t)}{d_2-d_1},
    \\
    & M(t) = c\left(\epsilon_1(t)-d_1\dfrac{\epsilon_2(t)-\epsilon_1(t)}{d_2-d_1}\right),
\end{align}\label{eq:reconstr_F}\end{subequations}
with $c$ constant linking the strain to the moment, obtained by calibration. 

Strain gauges are often affected by higher noise levels than other sensors, so it is common to equip the gauge amplifier with analogue filters; in the present setup, a low-pass filter with a cut-off frequency of $500$~Hz is installed in the amplifier to eliminate high-frequency noise from the signals and avoid aliasing. Moderately low frequencies are also affected by the filter, which introduces distortions predominantly in the form of frequency-dependent phase lags. Such lags can be easily quantified in advance, but their compensation in a RTHT would be cumbersome due to their frequency-dependent nature. If not properly compensated, a lag in the interface forces would introduce artificial damping in the hybrid structure, compromising the fidelity of the hybrid test, especially around the resonance peak. In contrast to the case of classical RTHT, with the proposed method the real-time signal is not actually used, but its harmonic coefficients are calculated asynchronously. Therefore, upon identification of the filter lag at each forcing frequency, the phase of the harmonic signal can be easily advanced with rotation matrices. If the test is run on a broad frequency range or if higher harmonics are included, different rotation matrices for each frequency and each harmonic could be used to advance the signal according to the identified phase lag. However, in the narrow operational frequency range of the test, which is $16$~Hz to $19$~Hz to cover the second resonance of the hybrid beam, the filter lag is identified to be almost constant and around $0.06$~rad.

Other than high-frequency noise, in the present setup, a prominent $50$~Hz (and its multiples) disturbance is generated by the lasers' power supply. This disturbance only affects the strain gauges signals as their cables are more exposed than others to the environment. Solutions to reduce $50$~Hz noise could be explored, such as shielding the lead wire or filtering the $50$~Hz signal during acquisition. However, in the context of the proposed iterative method,
%The choice of measuring forces from gauges is then possible only because, with the proposed iterative method, 
%the real-time signal is not actually used but only its harmonic coefficients are. Moreover, 
since the real-time constraint is lifted, the signals can be collected for several periods and then averaged before computing their harmonic coefficients. As long as the test frequency is far from $50$~Hz or its multiples, the contribution of power supply noise to the harmonic coefficients can be reduced by increasing the number of averaging periods to achieve the desired accuracy.
Let us assume that the actual shear force signal is a purely harmonic function with frequency $\Omega$ equal to the forcing frequency
\begin{equation}
    T(t) = \tilde{T}^{(C)} \cos(\Omega t) + \tilde{T}^{(S)} \sin(\Omega t).
\end{equation}
Then, the measured signal will be the summation of the actual force plus a noise contribution from the power supply as
\begin{equation}
    \tilde{T}(t) = T(t) + T_\mathcal{N}\sin(\rho_\mathcal{N}\Omega t+\phi_\mathcal{N})
\end{equation}
with $T_\mathcal{N}$ the amplitude of the noise, $\phi_\mathcal{N}$ its phase lag with respect to the real signal and $\rho_\mathcal{N}$ the ratio between the power supply noise frequency and that of the excitation. The errors on the harmonic coefficients of the measured force averaged over $n$ periods read
\begin{subequations}
\begin{align}
    &\tilde{T}^{(C)}-{T}^{(C)} = \frac{\Omega}{\pi n} \int_0^{\frac{2 n \pi}{\Omega}}T_\mathcal{N}\sin(\rho_\mathcal{N}\Omega t+\phi_\mathcal{N})\cos(\Omega t)dt
    =
    \frac{2 T_\mathcal{N}}{n \pi}\frac{\rho_\mathcal{N}}{\rho_\mathcal{N}^2-1}
    \sin(n \pi \rho_\mathcal{N} + \phi_\mathcal{N})
    \sin(  n \pi \rho_\mathcal{N})
    \\
    &\tilde{T}^{(S)}-{T}^{(S)} = \frac{\Omega}{\pi n} \int_0^{\frac{2 n \pi}{\Omega}}T_\mathcal{N}\sin(\rho_\mathcal{N}\Omega t+\phi_\mathcal{N})\sin(\Omega t)dt
    =
    \frac{2 T_\mathcal{N}}{n \pi}\frac{1}{\rho_\mathcal{N}^2-1}
    \cos(n \pi \rho_\mathcal{N} + \phi_\mathcal{N})
    \sin(  n \pi \rho_\mathcal{N})
\end{align}\label{eq:noise}
\end{subequations}

The closer the frequency of the disturbance to the forcing frequency --- or its multiples if the coefficients of higher harmonics are to be computed --- the higher its impact on the accuracy. Since the sine and cosine terms in the error equation are bounded, by increasing $n$ the desired accuracy for the shear force coefficients can be attained. This comes at the cost of increasing the waiting time, so the final choice of $n$ will be a trade-off between test speed and accuracy.

In the context of the present experiment, the excitation frequency is varied between $16$~Hz and $19$~Hz, so the influence of a $50$~Hz noise signal on the coefficients is mitigated by the frequency separation. However, the amplitude of the power supply noise is between $1$ and $10$ times higher than that of the actual signal for the range of vibration amplitude tested. A quantification of this disturbance in terms of displacement can be obtained by propagating it on the residual expressed in terms of displacement, which yields a value of $1.4$~mm disturbance. For this reason, the number of averaging periods $n$ is chosen equal to $30$ to ensure a maximum error in the harmonic coefficients of $0.013$~mm. As discussed in the next section, this value is related to the maximum accuracy that can be obtained by the hybrid test with this particular choice of averaging time.

Before discussing the results, we want to point out that, in the context of classical RTHT, measuring shear forces via strain gauges would not be a viable option. In fact, the use of analogue low-pass filters, necessary to avoid aliasing, introduces frequency-dependent lags, which are difficult to compensate in real-time. Moreover, the computation of forces through finite difference reduces the signal-to-noise ratio, which, for strain gauges, is already lower than for other sensors. On the other hand, reconstructing forces from force transducers could also introduce errors into the hybrid structure. There is no straightforward solution to measuring interface forces, especially in the case of continuous structures and when multiple degrees of freedom are controlled. In the context of the present method, however, forces reconstructed from strain gauges are undoubtedly more accurate than those measured at the transfer system, in that errors in the identification of the transfer system cannot be prevented, whereas both lags and noise in the gauge signals can be controlled. For this reason, although the transfer system is equipped with transducers, we rely on strain gauges to measure forces in the experiment. 

%manage/accomplish/perform/carry out

\section{Experimental results}
\label{sec:results}
The restrictive substructurability conditions of the structure render a classical RTHT challenging. Moreover, the presence of several frictional elements in the test rig, i.e., the rotating bearing and the slider, increase the difficulties in designing a control scheme in this case, due to the challenges in the identification of friction for a wide range of operating conditions. In contrast, the proposed method presented in \autoref{sec:prop} does not require any identification, provided that the forces are measured at the interface. In this section, the experimental results obtained with the proposed method on the hybrid cantilever described in \autoref{sec:beam} will be presented. 

We perform a frequency sweep in the range $16$~Hz to $19$~Hz on the hybrid beam presented in the previous section. The numerical model is obtained by discretisation of the PDE with the same properties used in~\autoref{sec:beam}, and length $L_N=170$~mm. The condensed dynamic stiffness matrix of the NS $D_N(\text{i}\Omega)$ is computed once for each frequency. The external forcing $F_N^{(e)}$ simply consists in a sinusoidal point force applied at the numerical side on the interface node. At each frequency, the proposed iterative method is used to find the voltage coefficients $V$ that zero the residual defined in~\autoref{eq:res}.

{A flowchart of the proposed method in relation to the present experiment is shown in~\autoref{fig:flowchart}. At the beginning of the iteration, the frequency and the Fourier coefficients of the demand $V$ are imposed on the actuators through an acquisition box based on the BeagleBone Black \cite{BartonBBB}. The same box also reads the signals measured by strain gauges and lasers in real time, reconstructs the interface quantities $u,\phi,T,M$ from~\autoref{eq:reconstr_U} and~\autoref{eq:reconstr_F}, and performs the computation of the instantaneous harmonic coefficients $F_P^{(i)}$ and $U_P$ over the selected waiting time of $n=30$ periods. The coefficients values are imported into Matlab where the residual is assembled following~\autoref{eq:res}. To ensure the system is at steady state, a check on the system's steadiness is done by computing the residual every 30 periods until the variation of its norm is lower than a certain \textit{transient tolerance}, chosen to be $0.013$~mm. Once the transient decays, the steady-state residual is used to compute the next guess on the voltage coefficients using Broyden's method~\cite{Broyden1965}. At each iteration, a new guess for the voltage coefficients is found and imposed on the actuators. The iterative procedure stops when the residual norm is below the tolerance chosen for convergence. If this tolerance is not met in $100$ iterations, the solution at that particular frequency is declared to be non-convergent. As discussed, a limitation on the choice of tolerance is given by the constant noise coming from the power supply that affects the residual norm up to $0.013$~mm. The chosen convergence tolerance is $0.02$~mm for each test in order to include other potential sources of noise other than the known source from the power supply. }

\begin{figure}[ht]
    \centering
    \begin{tikzpicture}[node distance=2.4cm]
    \node (start) [myblock] {Set parameters (frequency) and apply initial actuator demand $V$};
    \node (n1) [myblock,right of=start] {Measure sensors' signals for $n$ periods};
    \node (n2) [myblock,right of=n1] {Reconstruct interface quantities $u,\phi,T,M$ from \autoref{eq:reconstr_U} and \autoref{eq:reconstr_F}};
    \node (n3) [myblock,right of=n2] {Compute harmonic coefficients of interface quantities $U_P$, $F_P$};
    \node (n4) [myblock,right of=n3] {Assemble  residual (\autoref{eq:res}) every $n$ periods until steady};
    \node (nl) [myblock,right of=n4] {Is residual norm lower than tolerance?};
    \node (feedback) [feedback,above of=n3,yshift=-.1cm,xshift=0cm] {Update actuator demand $V$};
    \node (end) [myblock,right of=nl] {Save converged actuator demand $V$ and interface quantities};
    \draw [arrow] (start) -- (n1);
    \draw [arrow] (n1) -- (n2);
    \draw [arrow] (n2) -- (n3);
    \draw [arrow] (n3) -- (n4);
    \draw [arrow] (n4) -- (nl);
    \draw [arrow] (nl) -- node[anchor=south] {\scriptsize{Yes}} (end);
    \draw [arrow] (nl) |-  node[anchor=north west,yshift=-.6cm,xshift=-.1cm] {\scriptsize{No}} (feedback);
    \draw [arrow] (feedback) -| (n1);
    \end{tikzpicture}
    \caption{Flowchart of the iterative method for a single solution point}
    \label{fig:flowchart}
\end{figure}
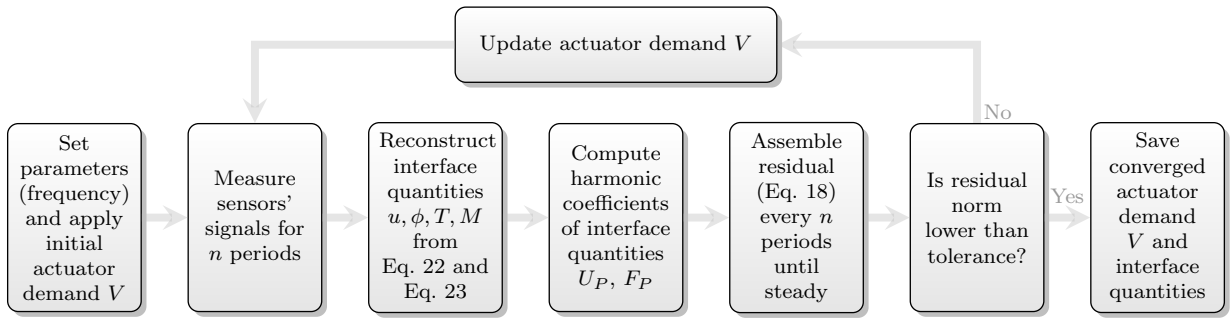

\subsection{Forced response tests}
\label{sec:frf}
As mentioned, one feature of PS in this case is the possibility of increasing its aerodynamic damping by changing the width of the plastic sheet placed on the beam. The damping ratio of the bare beam, identified from free decay vibrations, is lower than $1\%$ for the first two modes of the PS. Upon the addition of the plastic sheet, it was possible to increase the damping to approximately $2\%$ and thus perform the first set of tests under more robust conditions. 

The first round of tests is made up of six frequency sweeps on the hybrid cantilever in the most damped case, which corresponds to the largest width of the plastic sheet reported in \autoref{fig:cl}. Referring to \autoref{fig:exp1}, three forcing amplitudes have been used for the tests, two for each FRF: a low, a medium, and a high amplitude forcing which produced, in the most damped PS, a vibration amplitude of the interface section of around half, one, and two millimetres respectively. Since we targeted the resonance of the second mode of the HS, the displacement of the HS is composed mostly of this mode. Only one node is present in the second mode shape and is located in the PS portion of the HS, approximately $120$~mm from the tip, as expected. We can then estimate that a vibration of $1$~mm at the interface corresponds approximately to $2$~mm at the tip of the beam. 

\begin{figure}[ht]
    \centering
    \includegraphics[width=0.95\textwidth]{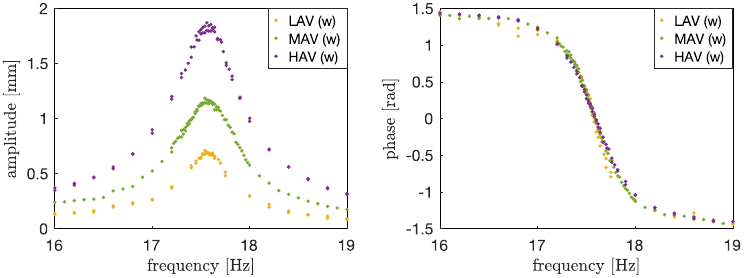}
    \caption{Experimental results obtained on the hybrid beam with the widest (w) plastic sheet for different excitation levels: low amplitude (LAV), medium amplitude (MAV), and high amplitude (HAV).}
    \label{fig:exp1}
\end{figure}

As shown in \autoref{fig:exp1}, the behaviour of the hybrid test is very robust, especially at medium amplitude vibration. In the peak of the purple curve, it is in fact possible to see a little discrepancy between two identical tests, probably due to the increasing effect of noise coming from the aerodynamics or due to changes in the environment at the time of the tests. On the other hand, in the phase plot, one can see that the yellow curve is aligned with the other two only around the resonance (zero degrees), whereas it departs from them far from resonance. This opposite behaviour is intuitively due to the fact that, as the tolerance of the iterations is constant for each experiment, a larger variability of the converged solution is allowed at very low amplitudes of vibration. However, all data points over a moderate range are generally very robust and demonstrate the repeatability of the test.

\subsection{Varying damping tests}
\label{sec:damping}
We now want to investigate how the robustness of the results obtained for the most damped case is affected by a decrease in aerodynamic damping. To do so, the second round of tests comprises six tests at different damping levels obtained by cutting the edges of the plastic sheet placed on the PS. Referring to \autoref{fig:exp2}, the green curve is the same as \autoref{fig:exp1}, which is the forced response obtained for the medium forcing level in the most damped case. We target this excitation level because it is the one that produced the cleanest results. The blue curve is obtained with a narrower size of the plastic, by cutting the sheet along the black lines reported in \autoref{fig:cl}, whereas the red curve corresponds to the bare beam case. To keep the vibration amplitude in a similar range, the blue curve is obtained with the same forcing level as the green curve, whereas the red curve is obtained with half the excitation. In this case, instead of plotting the amplitude of vibration, here we plot the amplitude scaled by the forcing level.

It is worth mentioning that every time a slice of plastic was cut off, it was placed on the top of the beam to keep the overall mass constant. It is possible to observe that as well as a decrease in the damping occurring for decreasing width, there is a small shift in the resonance peak towards higher frequencies although not as dramatic as the additional dissipation. This effect might be due to the fact that the air around the plastic sheet provided additional inertia to the system, or due to an increase in stiffness provided by the plastic sheets on the beam.

\begin{figure}[ht]
    \centering
    \includegraphics[width=0.95\textwidth]{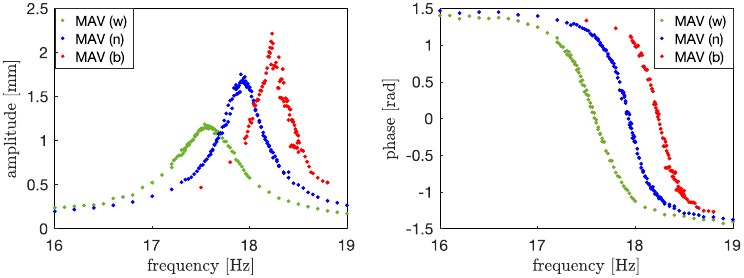}
    \caption{Experimental results obtained on the hybrid beam at medium excitation level (MAV) for different width of the plastic sheet on the beam: the widest one (w), the narrow one (n), and the bare beam case (b).}
    \label{fig:exp2}
\end{figure}

Six tests, two for each damping configuration, have been run by keeping the numerical model fixed because its update would have influenced the behaviour of the hybrid structure. In this way, the differences between each configuration are solely due to changes in the PS. From each response function, it is possible to estimate that the damping ratios for each case are approximately $0.5\%$, $1.0\%$, and $1.5\%$ for the green, blue, and red curves, respectively. This behaviour is consistent with the expectation based on the preliminary tests on each PS.

Moreover, we observe that the lower the damping in the PS, the less robust the results. An explanation for this behaviour can be found in the fact that environmental noise more significantly affects the coherence of the test when the damping is low, due to nonlinear effects. In addition to this, for lower damped structures, the transient takes a longer time to decay, and it can happen that some residual effects of the transient which are smaller than the tolerance are visible in the converged results. To quantify the loss of robustness with decreasing damping, in \autoref{fig:exp3} the number of iterations taken to reach convergence is reported. Broadly, the higher the damping, the faster the convergence. Moreover, around resonance the convergence in each of the tests is slower. This behaviour highlights how challenging the testing of lightly damped structures can be, as well as the importance of developing robust strategies to test these structures. 

\begin{figure}[ht]
    \centering
    \includegraphics[width=0.95\textwidth]{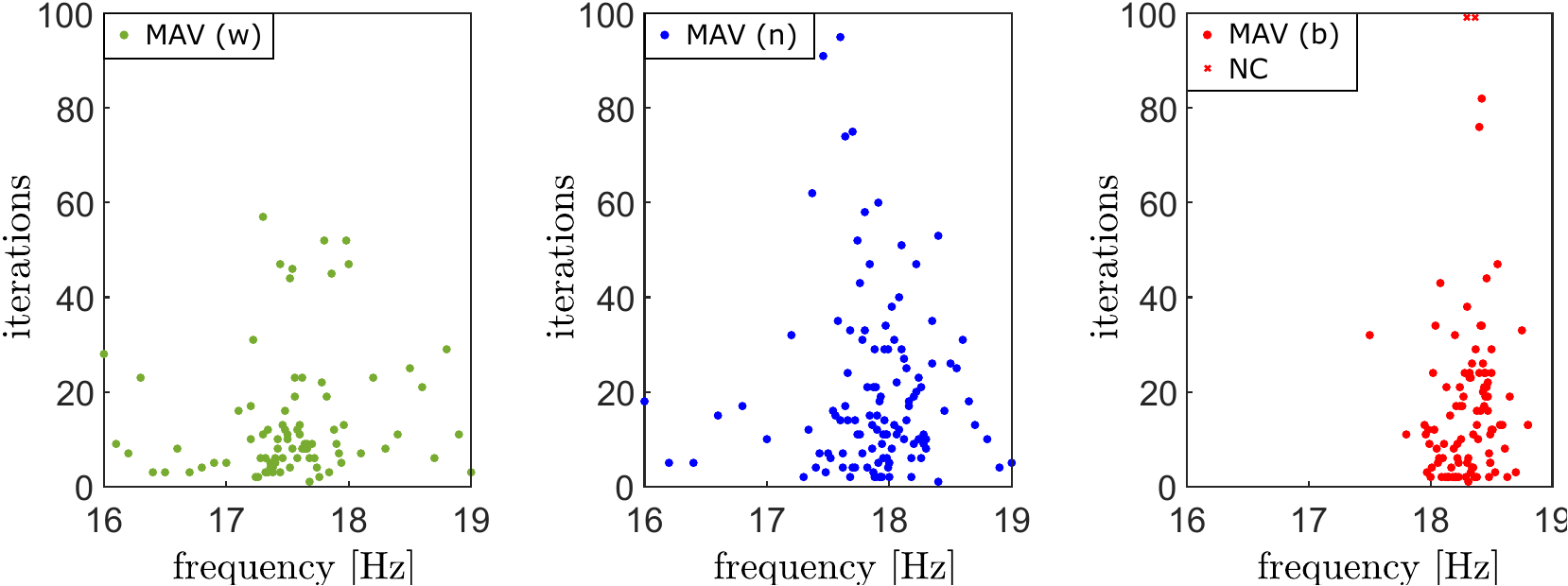}
    \caption{Number of iterations at medium excitation level (MAV) for different width of the plastic sheet on the beam: the widest one (w), the narrow one (n), and the bare beam case (b). Two solution points in the bare beam case reached the threshold of 100 iterations and have been labelled as non-converged (NC).}
    \label{fig:exp3}
\end{figure}

\subsection{Interface synchronisation}
\label{sec:interface}
To check the accuracy of the solution along the FRF, one can plot NS and PS quantities on the same plot or one versus the other for a given solution point. Here, we show the resonance peak of the bare beam case as an example. Very good interface synchronisation can be observed from \autoref{fig:harm}, on all interface variables. To give a better quantification of accuracy, the error in terms of time delay and relative amplification from PS to NS is given for the same quantities of \autoref{fig:harm} in \autoref{tab:harm}. The relative error is lower than $1\%$ on each variable which, for a solution point of around $2$~mm amplitude, is perfectly in line with the imposed tolerance of $0.02$~mm. Moreover, the maximum delay is on the rotation variable and is equal to a fifth of a millisecond, corresponding to a single time step at $5$~kHz sampling frequency. 
In terms of phase, the period here is $55$~ms long, so the phase lag equivalent is of $0.002$~rad. 

It can be seen that the displacement and rotation variables are slightly less accurate than the shear force and moment ones. The reason for this behaviour is that, in the tests, the residual is written in terms of forces rather than displacements. This could be seen as a force-controlled case, but in the framework of the present method neither displacement nor force is actually controlled and the different expression for the residual should be seen as a pure scaling law. Recalling \autoref{eq:com1}, both measured quantities appear in the residual and the only difference between a residual in terms of displacement and one in terms of force is in the premultiplication by the numerical dynamic stiffness matrix, that is, the above-mentioned scaling. 

The very good interface synchronisation shown here strongly suggests that the method is able to reproduce the behaviour of the original structure with a hybrid test, provided that the sensors perfectly measure the actual quantities at the interface. If that is not the case and, for instance, an error on the strain gauges is present, then a perfect synchronisation can still be attained, but that would not correspond to a complete reproduction of the original structure. To give an example, if the phase lag introduced by the amplifiers had not been corrected, the test would have produced a good match of the interface responses, but the FRF would appear much more damped than the one of \autoref{fig:exp2}. Obviously, these considerations also apply to a classical RTHT even though this problem is rarely considered in the RTHT literature, as sensor errors are negligible as compared to the interface errors. In this case, we could identify the PS beforehand, and by comparing the damping of the identification with that of the resulting FRF we are confident that the original structure is correctly reproduced. However, this is not possible in the general case where no information about the PS is available and the reason for the test is characterising the original structure. 
%If the behaviour at resonance is of interest, 
%Such considerations, must be take into account in the design stage of a RTHT otherwise one can trust results that do not replicate the original structure.
% the matching does not say much, because it can match but still be wrong! A phase lag in the forces for example could lead to a perfect match between PS and NS but the FRF will look much more damped than the real one. If the behaviour at resonance is the aim of the test, as it is in many engineering experiments, then the damping of the HS is the main actor: a fictitious damping due to sensor errors will completely mess up the results and the amplitude of oscillations. All the more true for lightly damped structures where a small fictitious damping can dramatically drop the resonance amplitude.
\begin{figure}[ht]
    \centering
    \includegraphics[width=0.85\textwidth]{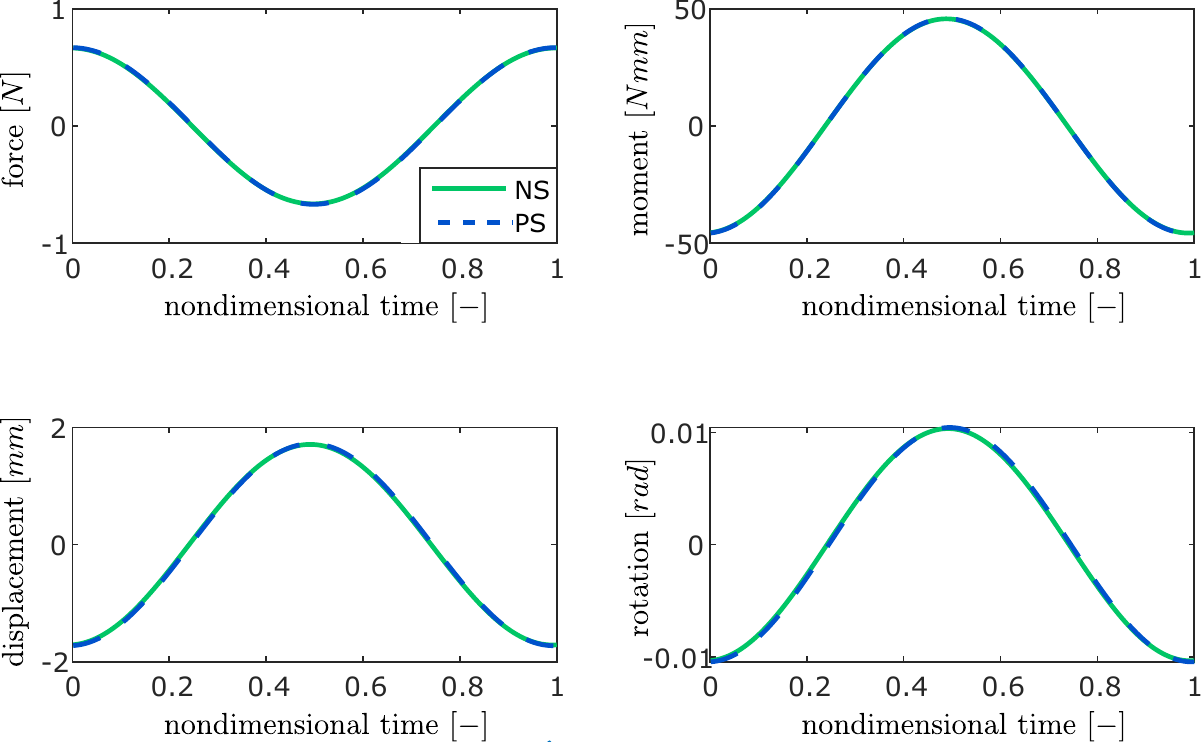}
    \caption{Solution at resonance for lower amplitude excitation of the bare beam case.}
    \label{fig:harm}
\end{figure}

\begin{table}[ht]
    \centering
    \begin{tabular}{lcccc}
    \toprule
         &Force & Moment & Displacement & Rotation  \\
         \midrule
         Delay of PS signal [ms]&
         -0.064 &-0.009&-0.15&-0.20
         \\
         Relative amplification of PS &0.52\% &0.31\%&0.72\%& 0.92\%\\
         \bottomrule
    \end{tabular}
    \caption{Errors of the converged solution depicted in Fig 9. Delay of PS signals with respect to NS signals and amplification of PS signals with respect to NS signals.}
    \label{tab:harm}
\end{table}

Another point is that the signals shown in \autoref{fig:harm} are the time reconstructions of the original signals from the computed harmonic coefficients. In fact, the method can only control the harmonic behaviour of the structure and cannot control any non-periodicity in the signal or the harmonics that are not included. Before starting the acquisition of time signals, to ensure that the transient is decayed, the steadiness of the harmonic coefficient is checked against a defined \textit{transient tolerance}. The non-periodicity of the signal can then be controlled thanks to this strategy. As per the higher harmonics not included in the discretisation, they will not register in this test due to their orthogonality to the lower ones. Their presence must therefore be checked separately, for example by computing their values and assessing whether the inclusion of higher harmonics is necessary or if they can be neglected. In \autoref{fig:harm2}, the harmonic reconstruction of the signal and the time signal averaged over 30 periods are shown. A small contribution from the second harmonic is present in the force and moment signals. Such a contribution cannot be seen in the displacement and rotation, so they have not been reported here. The most likely explanation for this observation is that some nonlinearities in the actuation and transfer system generate higher harmonics that are amplified by the PS. In fact, it is computed that the first clamped mode of the PS is approximately twice the operational frequency chosen, so the first \textit{clamped interface} mode of the PS discussed in \autoref{sec:beam} generates high response forces from small interface displacements in the PS. However, since the NS is a linear structure and the PS behaviour could also be treated as linear, this second-harmonic response is fully orthogonal to the first-harmonic one and thus should not affect the response of the hybrid structure. That is to say, if the second harmonic were included in the test to suppress this response, no difference in the forced response results would have been observed. For this reason, we chose not to perform further runs including the higher harmonics. 

{Overall, the experiment was successful especially given the challenging settings of low damping and two controlled degrees of freedom. However, the difficulty in accurately measuring interface forces limits the capabilities of the test rig, which would require further refinement to be used on a wider range of structures, e.g. nonlinear beams.
%The capabilities of the test rig are, however, still limited, mainly due to difficulty of measuring interface forces accurately. Further developments of the test rig would be then needed to extend  However, the hybrid testing of a wider range of structures would require the inclusion of higher harmonics, therefore further developments of the test rig would be necessary especially in terms of isolation from environmental noise and measurement of interface forces. 
}
%First of all, the reduction of environmental noise would enable (i) faster experiment that require less periods for the averaging or (i) the treatment of nonlinear structures for which higher harmonics are needed, which could go too close to the 50 Hz. Another possibility could be to avoid using SG and only rely on the load cells but that would require a redesign of the transfer system so to measure the interface force after the bearings rather than before, so the identification of the transfer system is possible. }
%This mode high wavelength so amplifies T and M?

% we could eliminate this but both NS and PS are linear so we are confident this does not affect the results
% it'll be slow in the present setup 180 periods to make it work
% we could shield the cables

\begin{figure}[ht]
    \centering
    \includegraphics[width=.92\textwidth]{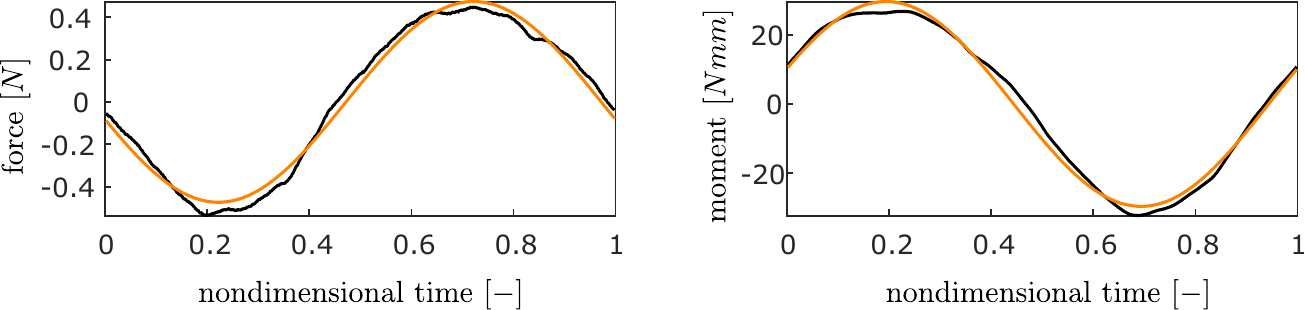}
    \caption{Solution after resonance. Averaged signal and signal from harmonic reconstruction.}
    \label{fig:harm2}
\end{figure}

\section{Conclusions}
Real-time hybrid testing (RTHT) allows testing critical components of large structures within the context of the whole assembly. To accurately reproduce the behaviour of the original structure, synchronisation of forces and displacements at the interface is required. However, interface delays and distortion introduced by the control system can cause instability and poor accuracy in the response of the hybrid structure, thus invalidating the purpose of the test. A typical mitigating strategy is to use delay compensation techniques, but the implementation of these strategies is often complex and rig dependent. 

In this paper we propose an iterative method, suitable for testing mechanical components at resonance. The method, detailed in \autoref{sec:prop}, handles the real-time constraint by enforcing interface synchronisation in the Fourier domain, which is achieved through a solution of algebraic equations using quasi-Newton methods. The approach in the presented form is only suitable for assessing the periodic steady state of structures but has the great advantage of being rig and structure independent. In fact, with the proposed method, the accuracy of the response of the hybrid structure in reproducing the response of the original structure is only limited by sensing errors. 

To mitigate reconstruction errors arising from force sensors placed far from the interface, in this work, we also introduce a novel methodology for directly measuring forces at the interface. In \autoref{sec:gauges} it is shown that by placing two strain gauges near the clamp of the physical substructure the interface bending moment and shear force can be reconstructed by extrapolation and finite difference, respectively. This sensing strategy has proven to be more accurate than using force transducers, but it is only viable in the context of the proposed method, whereby the noise can be reduced up to a required accuracy by averaging the strain gauge signals over multiple periods during each iteration step. 

We applied the iterative method proposed in this work to the hybrid testing of a cantilever beam, a first step towards enabling the hybrid testing of aircraft wings, where numerical models are reliable at the root but physical experiments might be needed at the tip. The feasibility of the test is investigated by focusing on the bending motion of a two-degree-of-freedom cantilever beam at the interface. 

In \autoref{sec:substruct}, we numerically assess the substructurability of this structure, demonstrating that classical RTHT is particularly challenging in this case due to potential instability caused by delays as small as $1$~ms irrespective of the position of the interface along the beam.

The experimental results obtained with the proposed method are presented in \autoref{sec:results}, where the robust behaviour of the hybrid test, particularly at medium-amplitude vibration, is shown. %We notice that the convergence of the iterations is slower around resonance. 
Additionally, we observed that the results are less robust with lower damping in the physical substructure, which we ascribe to the fact that environmental noise has a more significant impact on test coherence under low damping conditions. Nevertheless, the hybrid test can effectively replicate the behaviour of the original structure with excellent interface synchronisation, corresponding to a single time step at $5$~kHz, which would be extremely challenging to obtain with classical RTHT.

%\clearpage
\section*{Conflict of interests}
The authors declare that they have no known competing interests.
\section*{Acknowledgements}
The authors gratefully acknowledge the support of the Engineering and Physical Sciences Research Council,
United Kingdom through the award of a Programme Grant ``Digital Twins for Improved Dynamic Design'', Grant
Number EP/R006768.
\section*{Data availability}
{The implementation of the proposed method and all the experimental data presented in this paper are available from \cite{DataRepo}} or via the University of Bristol Data Repository at \url{https://data.bris.ac.uk/data/}}

\bibliographystyle{elsarticle-num} 
\bibliography{biblio}

% %% The Appendices part is started with the command \appendix;
% %% appendix sections are then done as normal sections
% \appendix
% \section{Eigenvalues Sensitivity}\label{sec:app}
% Displacement control:
% \begin{equation}
%     X_{Pb} = e^{-\lambda\tau} X_{Nb}
% \end{equation}

% Displacement control:
% \begin{equation}
% D_\text{displ}=
% \begin{bmatrix}
%     D_{Nii} &D_{Nib}e^{+\lambda\tau} &       0       \\
%     D_{Nbi} &D_{Nbb}e^{+\lambda\tau}+D_{Pbb} &D_{Pbi}\\
%         0   &D_{Pib}         &D_{Pii}
% \end{bmatrix}
% \end{equation}

% Variation of the eigenvalue of $D_0$ due to the variation of a parameter $\gamma$:
% \begin{equation}
%     \lambda_\gamma = -\dfrac{\phi^{T}D_\gamma\phi}{\phi^{T}D_\lambda\phi}
% \end{equation}

% with:
% \begin{equation}
%     D_\lambda = 2\lambda M + C
% \end{equation}
% and:
% \begin{equation}
%     D_\gamma = \lambda^2 M_\gamma + \lambda C_\gamma + K_\gamma
% \end{equation}

% If the error is an amplification error on the displacement:
% \begin{equation}
% D_{\gamma}=
% \begin{bmatrix}
%     0 &D_{Nib} &       0       \\
%     0 &D_{Nbb} &       0\\
%         0      &       0         &0
% \end{bmatrix}
% \end{equation}
% If it is a lag then it is the same multiplied by $i$, if a delay by $\lambda$.


\end{document}